\documentclass[aps,prb,reprint,showpacs,floatfix]{revtex4-1}
\usepackage{amsmath,graphics,epsfig,color,verbatim,amsfonts}

\let\oldhat\hat

\renewcommand{\hat}[1]{\oldhat{\mathbf{#1}}}

\begin{document}

\title{Compositional phase stability of strongly correlated electron materials within DFT+$U$}
\author{Eric B. Isaacs}
\email{eric.isaacs@columbia.edu}
\author{Chris A. Marianetti}
\email{chris.marianetti@columbia.edu}
\affiliation{Department of Applied Physics and Applied Mathematics, Columbia University, New York, NY 10027}

\begin{abstract}
Predicting the compositional phase stability of strongly correlated
electron materials is an outstanding challenge in condensed matter
physics. In this work, we employ the density functional theory plus
$U$ (DFT+$U$) formalism to address the effects of local correlations
due to transition metal $d$ electrons on compositional phase stability
in the prototype phase stable and separating materials Li$_x$CoO$_2$
and olivine Li$_x$FePO$_4$, respectively. We exploit a new spectral
decomposition of the DFT+$U$ total energy, revealing the distinct
roles of the filling and ordering of the $d$ orbital correlated
subspace. The on-site interaction $U$ drives both of these very
different materials systems towards phase separation, stemming from
enhanced ordering of the $d$ orbital occupancies in the $x=0$ and
$x=1$ species, whereas changes in the overall filling of the $d$ shell
contribute negligibly. We show that DFT+$U$ formation energies are
qualitatively consistent with experiments for phase stable
Li$_x$CoO$_2$, phase separating Li$_x$FePO$_4$, and phase stable
Li$_x$CoPO$_4$. However, we find that charge ordering plays a critical
role in the energetics at intermediate $x$, strongly dampening the
tendency for the Hubbard $U$ to drive phase separation. Most
relevantly, the phase stability of Li$_{1/2}$CoO$_2$ within DFT+$U$ is
qualitatively incorrect without allowing charge ordering, which is
problematic given that neither charge ordering nor the band gap that
it induces are observed in experiment. We demonstrate that charge
ordering arises from the correlated subspace interaction energy as
opposed to the double counting. Additionally, we predict the Li
order-disorder transition temperature for Li$_{1/2}$CoO$_2$,
demonstrating that the unphysical charge ordering within DFT+$U$
renders the method problematic, often producing unrealistically large
results. Our findings motivate the need for other advanced techniques,
such as DFT plus dynamical mean-field theory, for total energies in
strongly correlated materials.
\end{abstract}

\date{\today}
\pacs{71.15.Mb, 71.27.+a, 81.30.Bx, 82.47.Aa}
\maketitle

\section{Introduction}

Strongly correlated materials (SCMs), in which density functional
theory
(DFT)\cite{hohenberg_inhomogeneous_1964,kohn_self-consistent_1965}
calculations break down for selected observables due to strong
electron-electron interactions, are at the forefront of condensed
matter physics.\cite{kotliar_strongly_2004,morosan_strongly_2012}
Phenomenologically, SCMs exist in a ground state which is in the
vicinity of a Mott transition
\cite{mott_metal-insulator_1968,imada_metal-insulator_1998} whereby
electronic hopping may be overwhelmed by local interactions, resulting
in an insulating state. Realizations of SCMs often contain atoms with
open-shell $d$ or $f$ electrons, such as the high-temperature
superconducting cuprates\cite{bednorz_possible_1986}, colossal
magnetoresistance manganites\cite{ramirez_colossal_1997}, and heavy
fermion actinide based materials.\cite{coleman_heavy_2007}

Predicting the properties of strongly correlated materials is an
outstanding problem in solid state physics. The standard approach of
DFT, which is the most generic theory of electronic structure for
materials physics, is in principle an exact theory for the ground
state energy and electron density of any many-electron system.
However, the exact exchange-correlation functional is unknown and must
be approximated; and in practice DFT is often unable to capture
critical aspects of the physics of strongly correlated
materials.\cite{jones_density_1989,jones_density_2015,kotliar_strongly_2004}
The difficulty of constructing increasingly intelligent
exchange-correlation density functionals led to pursuit of additional
variables that could be more easily approximated; one such variable is
the local Green function for a set of correlated orbitals in the
material (e.g $d$ orbitals)\cite{kotliar_electronic_2006}. A
corresponding approximation for this density plus local Green
functional theory leads to DFT plus dynamical mean-field theory
(DFT+DMFT). A further Hartree approximation to the DMFT impurity
problem leads to the more ubiquitous DFT+$U$ approach, which can be
counterintuitive given that the latter preceded the former
historically\cite{anisimov_first-principles_1997,kotliar_electronic_2006}.
The long-term goal of these techniques is their application in an
unbiased manner to any material containing localized orbitals,
nominally or otherwise. Compositional phase stability is a particular
stringent test since it relies on total energies from multiple
compositions, making it more difficult to cancel errors by varying the
few still maturing aspects of the formalism.

One of many important contexts highlighting the need for accurate
total energy methods for strongly correlated materials is that of
doped transition metal oxides, which include doped Mott insulators.
From a technological perspective, Li-based transition metal oxides
form the basis of rechargeable battery cathodes. Such cathodes
typically employ transition metals and oxygen or oxygen-containing
anion groups and, since they have open $d$ electron shells, are
susceptible to strong correlation physics. One fundamental
characteristic of a cathode material is whether there is a stable
solid solution for intermediate Li concentration ($0<x<1$), as
dictated by the formation energy of the compound. The formation energy
has a strong impact on the charge/discharge mechanism and also has
implications for the voltage and capacity of the battery.

\begin{figure}[htbp]
\begin{center}
\includegraphics[width=\linewidth]{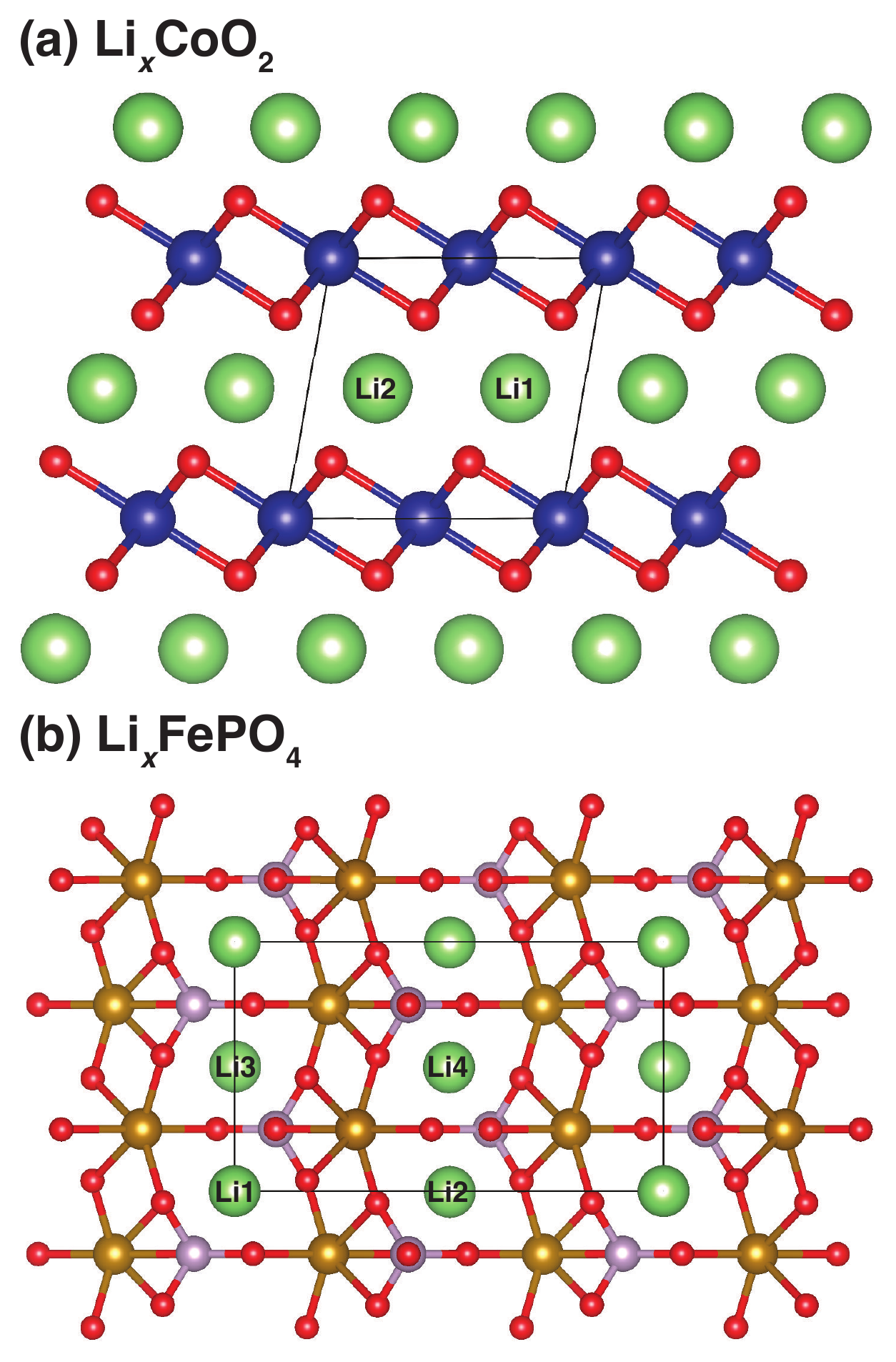}
\end{center}
\caption{Crystal structures of (a) Li$_x$CoO$_2$ and (b)
  Li$_x$FePO$_4$. The large green, medium blue, medium gold, small
  purple, and small red spheres represent ionic positions of Li, Co,
  Fe, P, and O, respectively. The black lines indicate for (a) the
  unit cell of the lowest-energy Li$_{1/2}$CoO$_2$ configuration and
  for (b) the primitive unit cell. Images of crystal structures are
  generated using
  \textsc{vesta}.\cite{momma2011vesta}\label{structures}}
\end{figure}

Many of the current cathode materials utilized today are based on the
material Li$_x$CoO$_2$, shown in Fig. \ref{structures}(a) for $x=1$,
which can store Li ions in between its CoO$_2$
layers.\cite{mizushima_lixcoo2_1980} Li$_x$CoO$_2$ has several stable
intermediate
phases.\cite{reimers_electrochemical_1992,van_der_ven_first-principles_1998,wolverton_first-principles_1998}
In contrast, olivine Li$_x$FePO$_4$
\cite{padhi_phospho-olivines_1997,padhi_effect_1997} [see Fig.
  \ref{structures}(b)] does not have any stable compounds for
intermediate $x$. It instead exhibits phase separation into its fully
lithiated ($x=0$) and fully delithiated ($x=1$) endmembers, which may
contribute to slow charge/discharge kinetics.
\cite{padhi_phospho-olivines_1997,delacourt_existence_2005,dodd_phase_2006}

Previous work demonstrates that DFT predicts very stable compounds of
intermediate $x$ in Li$_x$FePO$_4$, in stark disagreement with
experiment.\cite{zhou_phase_2004} This is a flagrant failure of the
DFT method. In contrast, DFT does properly capture the phase stability
of
Li$_x$CoO$_2$.\cite{van_der_ven_first-principles_1998,wolverton_first-principles_1998}
A beyond-DFT approach combining DFT and Hartree-Fock exchange, called
a hybrid functional, also fails to capture the total phase separation
in Li$_x$FePO$_4$.\cite{ong_comparison_2011} DFT+$U$ calculations,
which include an explicit on-site Coulomb interaction $U$ for the
transition metal $d$ electrons, were found to predict phase separation
in Li$_x$FePO$_4$ for sufficiently high values of
$U$.\cite{zhou_phase_2004} In later work, a cluster expansion based on
such DFT+$U$ energetics rationalized this phase separation in terms of
strong Li--electron interactions.\cite{zhou_configurational_2006} A
lingering question, which we will address in this work, is whether
DFT+$U$ can be applied to Li$_x$CoO$_2$
\cite{andriyevsky_electronic_2014,aykol_local_2014,aykol_van_2015,seo_calibrating_2015}
without degrading the already reasonable DFT results, such that
DFT+$U$ can give a robust description for both of these prototype
phase stable and phase separating systems.

The DFT+$U$ study for Li$_x$FePO$_4$ claims that the instability of
compounds of intermediate $x$ relates to energy penalties stemming
from charge ordering (CO), a symmetry breaking in which the number of
electrons on different transition metal sites
differs.\cite{zhou_phase_2004} However, it is unclear if the CO is
physical since static ordering is the only way in which a static
mean-field theory like DFT+$U$ can mimic strong electronic
correlations. Furthermore, the phase separation in Li$_x$FePO$_4$ is
fundamentally puzzling since with reasonable parameters the canonical
model Hamiltonians describing strong correlations such as the Hubbard
and $t$--$J$ models do not exhibit phase separation for extended
regions of the phase
diagram.\cite{visscher_phase_1974,marder_phase_1990}

To explore such issues, in this work we employ extensive DFT+$U$
calculations to understand in detail the impact of electronic
correlations on phase stability in correlated intercalation materials.
We focus on phase stable Li$_x$CoO$_2$ and phase separating
Li$_x$FePO$_4$ and explicitly investigate the role of CO and
structural relaxations. A new energy decomposition is elucidated to
quantitatively analyze the impact of $U$ on the overall filling and on
the ordering of orbitals within the $d$ shell. Comparison is made to
the Li$_x$CoPO$_4$ system, which is isostructural to Li$_x$FePO$_4$
but does have a stable compound of intermediate $x$. We also
investigate another physical observable, the Li order-disorder
transition temperature, for Li$_{1/2}$CoO$_2$ to provide another clear
benchmark of DFT+$U$ theory. The tendency for CO in DFT+$U$, how it
impacts these thermodynamic properties, and whether it is physical or
not is investigated and discussed. A thorough understanding at the
DFT+$U$ level of theory is critical as we advance to more
sophisticated methods for studying phase stability in strongly
correlated materials.

For all the materials we have investigated, we find that $U$
destabilizes compositions at intermediate $x$. This results because
$U$ most strongly affects the endmembers by enhancing the deviations
in the $d$ orbital occupancies from their average value; compounds of
intermediate $x$ experience the same effect but are unable to
orbitally polarize to the same extent, which directly drives them
towards phase separation as $U$ increases. CO and structural
relaxations serve to dampen, but not eliminate, this fundamental
effect. However, CO leads to the formation of a band gap for
Li$_{1/2}$CoO$_2$ that is not observed experimentally, suggesting it
is an unphysical artifact of DFT+$U$. Furthermore, CO leads to highly
erratic estimations of the Li order-disorder transition temperature
for Li$_{1/2}$CoO$_2$, demonstrating a serious shortcoming of the
DFT+$U$ method. Our calculations demonstrate that the interaction term
rather than the double counting term in DFT+$U$ is responsible for the
CO, which suggests that more accurate approaches such as DFT plus
dynamical mean-field theory (DFT+DMFT) may be necessary to accurately
describe this class of materials.

The rest of this paper is organized as follows. Section
\ref{methodology} discusses the DFT+$U$ approach to materials with
strong electronic correlations and elucidates our new energy
decomposition in this framework. Section \ref{computational_details}
describes the computational details for the simulations performed in
this work. The electronic structure of the endmembers of Li$_x$CoO$_2$
and Li$_x$FePO$_4$ within DFT are described in Sec. \ref{endmembers}.
The impact of $U$ on the electronic structure of Li$_x$CoO$_2$ is
discussed in Sec. \ref{lixcoo2_electronic_structure}. The origin of
the tendency for CO in DFT+$U$, taking Li$_x$CoO$_2$ as an example, is
the focus of Sec. \ref{charge_ordering}. The impact of $U$ on the
phase stability of Li$_x$CoO$_2$ is discussed in Sec.
\ref{lixcoo2_phase_stability}. The electronic structure and phase
stability of Li$_x$FePO$_4$ within DFT+$U$ are described in Sec.
\ref{lixfepo4_electronic_structure} and Sec.
\ref{lixfepo4_phase_stability}, respectively. Section \ref{lixcopo4}
discusses the formation energy trends for Li$_x$CoPO$_4$. Section
\ref{strawman} elucidates the role of the double counting on the
formation energy trends. The average intercalation voltages for
Li$_x$CoO$_2$ and Li$_x$FePO$_4$ are presented in Sec. \ref{voltage}.
Section \ref{od_temp} focuses on the Li order-disorder transition
temperature of Li$_{1/2}$CoO$_2$ and its dependence on CO. Finally,
conclusions are presented in Sec. \ref{conclusions}.

\section{Methodology}\label{methodology}

\subsection{DFT+$U$ approach for correlated materials}

The idea of DFT+$U$\cite{himmetoglu_hubbard-corrected_2014} is to
provide an improved treatment of electronic correlations by using not
only the density $\rho$ as a primary variable of the energy
functional, but also the single-particle density matrix of a relevant
set of local orbitals associated with strong correlations. This set of
orbitals, which form the correlated subspace, are typically localized
electronic states having $d$ or $f$ character and in practice are
defined using Wannier functions or atomic orbitals
$|\phi_m^\tau\rangle$ labeled by ionic site $\tau$ and angular
momentum projection $m$.

Having defined the correlated subspace, one needs to construct an
approximation for the energy as a functional of the density and the
single-particle density matrix of the correlated subspace. This is
typically approximated using two separate additive functionals:
\begin{equation}E_{\mathrm{DFT}+U}[\rho,n^{\tau s}]=E_{\mathrm{DFT}}[\rho]+E_U[n^{\tau s}]-E_{\mathrm{dc}}[n^{\tau s}].\end{equation}
where $n^{\tau s}$ is the local single-particle density matrix for
spin projection $s$ and $E_{\mathrm{DFT}}[\rho]$ is the usual
Kohn-Sham DFT energy functional using one of the many possible
approximations to the exchange-correlation energy such as the local
density approximation
(LDA)\cite{ceperley_ground_1980,vosko_accurate_1980,perdew_self-interaction_1981,perdew_accurate_1992}
or the generalized gradient approximation
(GGA).\cite{perdew_atoms_1992,perdew_generalized_1996} The functional
of the density matrix $E_U[n^{\tau s}]$, to be defined mathematically
below, is given by the Hartree-Fock interaction energy based upon a
set of screened interactions within the correlated subspace. There is
a clear double counting (dc) problem with such a formulation, as the
LDA or GGA already accounts for some interactions of the density
arising from the correlated subspace, and therefore a double-counting
energy $E_{\mathrm{dc}}[n^{\tau s}]$ must be defined and subtracted.


It is common to employ the spin-dependent formulation of DFT (SDFT)
rather than pure DFT, in which case the total energy expression
becomes \begin{align}E_{\mathrm{SDFT}+U}[\rho^s,n^{\tau
      s}]=&E_{\mathrm{SDFT}}[\rho^s]\nonumber\\ &+E_U[n^{\tau
      s}]-E_{\mathrm{dc}}[n^{\tau s}],\end{align} where $\rho^s$ is
the density for spin projection $s$. In the simplified
rotationally-invariant formalism of Dudarev \textit{et
  al.}\cite{dudarev_electron-energy-loss_1998} the interaction term,
which we always write in the diagonal eigenbasis of the density
matrix, is \begin{equation}E_U[n^{\tau
      s}]=\frac{1}{2}U\sum_{\tau,ms\neq m's'}n_m^{\tau s}n_{m'}^{\tau
    s'},\label{dftu_u_eqn}\end{equation} where $n_m^{\tau s}$ is the
$m$th eigenvalue of $n^{\tau s}$ and $U$ represents the screened
on-site Coulomb energy between different spin-orbitals within the
correlated subspace. This approach is equivalent to the full
rotationally-invariant formalism by Liechtenstein \textit{et
  al.}\cite{liechtenstein_density-functional_1995} if the on-site
exchange parameter $J$ is set to 0, which can be justified by previous
work indicating that SDFT already contains an intrinsic
$J$.\cite{chen_density_2015,park_density_2015} The most commonly used
dc is the fully-localized-limit (FLL)
type \begin{equation}E_{\mathrm{dc}}[n^{\tau
      s}]=\frac{1}{2}{U}\sum_{\tau}N^{\tau}(N^{\tau}-1),\label{flldc}\end{equation}
in which $N^{\tau}=\sum_{m,s}n_{m}^{\tau s}$ is the total correlated
occupancy on a site. In this work we focus on materials in which $d$
states form the correlated subspace, so at times we refer to this
quantity as $N_d^{\tau}$ (or $N_d$ if there is only a single site).
This double counting energy is equal to $E_U[n^{\tau s}]$ in the limit
in which all $n_{m}^{\tau s}$ take on values of 0 or 1.

Using this dc form, the total energy expression can be rewritten
as \begin{align}E_{\mathrm{SDFT}+U}[\rho^s,n^{\tau
      s}]=&E_{\mathrm{SDFT}}[\rho^s]\nonumber\\ &+\frac{1}{2}U\sum_{\tau,m,s}n_{m}^{\tau
    s}(1-n_{m}^{\tau s}).\end{align} This form illustrates that
DFT+$U$ penalizes fractional occupancy of the correlated orbitals and
serves to restore the derivative discontinuity of the total energy
with respect to total particle number that is missing in approximate
DFT.\cite{perdew_density-functional_1982,anisimov_density-functional_1993,himmetoglu_hubbard-corrected_2014,cococcioni_linear_2005}

It is important to emphasize the connection between DFT+$U$ and
DFT+DMFT to in order to appreciate the limitations of the DFT+$U$
approach.\cite{kotliar_electronic_2006} DFT+DMFT corresponds to a
functional of the density and the local Green function of the
correlated orbitals, which contains the single-particle density matrix
in addition to information about single-particle excitations. This
functional is approximated using, in part, DMFT, which requires the
solution of the DMFT impurity problem. Since the DMFT impurity problem
itself is a difficult (though tractable) many-body problem, one
approach is to make further approximations and solve the DMFT impurity
problem within the Hartree-Fock approximation, which results in the
DFT+$U$ approach. Alternatively, it may be possible to exactly solve
the DMFT impurity problem using quantum Monte Carlo (QMC)
techniques\cite{gull_continuous-time_2011,haule_quantum_2007}, which
can have a profound influence on the energetics as compared to
DFT+$U$\cite{park_total_2014,park_computing_2014,chen_density_2015}.
In particular, the error associated with DFT+$U$ was shown to strongly
depend on the spin state of Fe in a spin crossover
molecule\cite{chen_density_2015}, which may suggest that DFT+$U$
errors will strongly depend on Li concentration in cathode materials
and therefore affect phase stability and battery voltages. The quality
of DFT+DMFT (solved using QMC) is superior to DFT+$U$ in every sense,
but the former is far more computationally expensive and complex. This
current study will give us a detailed understanding of how DFT+$U$
performs in the context of phase stability, sometimes exposing obvious
deficiencies but also demonstrating enhancements over DFT, and provide
a baseline on which DFT+DMFT may improve. A forthcoming publication
will then extend the current study using DFT+DMFT.

\subsection{Filling and ordering energy decomposition in DFT+$U$}\label{decompositioneqns}

Two important aspects of DFT+$U$ calculations in correlated materials
are (1) the total number of correlated ($d$ or $f$) electrons per site
and (2) the ordering of electrons within the correlated subspace based
on spin (magnetism), angular momentum (orbital ordering), or site
(CO). These two effects can be labeled as the ``filling'' and
``ordering'' of the correlated orbitals, respectively. Here we
elucidate a decomposition of the interaction and dc energies that
enables a useful decoupling of these two effects for analysis of
DFT+$U$ results. This spectral decomposition was also very briefly
introduced in our previous work.\cite{isaacs_electronic_2016}

To represent the average filling on a site we define the orbital
occupancy mean \begin{equation}\mu_{\tau}=\frac{\sum_{m,s}n_{m}^{\tau
      s}}{N_{\mathrm{orb}}},\end{equation} where
$N_{\mathrm{orb}}=\sum_{m,s}1$ is the number of correlated
spin-orbitals per site (e.g. 10 for the $d$ shell). We then rewrite
the interaction and dc energies to contain only terms containing the
mean $\mu_{\tau}$ or the deviation from the mean $n_{m}^{\tau
  s}-\mu_{\tau}$.

\begin{align}
E_U-E_{\mathrm{dc}}=\frac{1}{2}U&\sum_{\tau,m,s}\left[n_{m}^{\tau s}-(n_{m}^{\tau s})^2\right]\nonumber\\
=\frac{1}{2}U&\sum_{\tau,m,s}\left\{ n_{m}^{\tau s}-\left[(n_{m}^{\tau s}-\mu_\tau)^2-\mu_\tau^2+2n_{m}^{\tau s}\mu_\tau\right]\right\}\nonumber\\
=\frac{1}{2}U&\sum_{\tau}\bigg[  \sum_{m,s}n_{m}^{\tau s} -\sum_{m,s} (n_{m}^{\tau s}-\mu_\tau)^2 +\nonumber\\ &\hspace{9mm}\mu_\tau^2\sum_{m,s}1 - 2\mu_\tau\sum_{m,s}n_{m}^{\tau s} \bigg]\nonumber\\
=\frac{1}{2}U&N_{\mathrm{orb}}\sum_{\tau}\left[ \mu_\tau -\mu_{\tau}^2 -\frac{\sum_{m,s} (n_{m}^{\tau s}-\mu_\tau)^2  }{N_{\mathrm{orb}}} \right]\nonumber\\
=\frac{1}{2}U&N_{\mathrm{orb}}\sum_{\tau}\left[ \mu_{\tau}(1-\mu_{\tau}) -\sigma_\tau^2 \right],
\end{align}

where \begin{equation}\sigma_{\tau}=\sqrt{\frac{\sum_{m,s}(n_{m}^{\tau
        s}-\mu_{\tau})^2}{N_{\mathrm{orb}}}}\end{equation} is the
standard deviation of the occupancies for site $\tau$. We call
$\mu_{\tau}(1-\mu_{\tau})$ the filling factor and $\sigma_{\tau}^2$
the ordering factor, and at times in our discussion of results we drop
the $\tau$ subscript on $\mu_{\tau}$ and $\sigma_{\tau}$ for
convenience. $E_U-E_{\mathrm{dc}}$ is thus rewritten as
$E_{\mathrm{fill}}+E_{\mathrm{ord}}$
with: \begin{equation}E_{\mathrm{fill}}=\frac{1}{2}UN_{\mathrm{orb}}\sum_{\tau}\mu_{\tau}(1-\mu_{\tau})\end{equation}
and \begin{equation}E_{\mathrm{ord}}=-\frac{1}{2}UN_{\mathrm{orb}}\sum_{\tau}\sigma_{\tau}^2.\label{eord}\end{equation}

These terms encapsulate the two ways in which a system can avoid
paying the Coulomb energetic penalty $U$. From the filling term, the
system can minimize $\mu_{\tau}(1-\mu_{\tau})$ by moving towards
completely empty ($\mu_{\tau}=0$) or completely filled
($\mu_{\tau}=1$) correlated shells on average. From the ordering term,
the system can maximize $\sigma_{\tau}^2$ by enhancing the spread in
$n^{\tau s}_m$ via some type of ordering. Note that $n_m^{\tau s}$
values for nominally-unoccupied orbitals still contribute to
$\sigma_{\tau}$, so this quantity is distinct from other measures of
orbital ordering that are determined from low-energy orbitals (e.g.
Wannier orbitals defined from a small energy window). One can rewrite
$E_U$ and $E_{\mathrm{dc}}$ in terms of $\mu_{\tau}$ and
$\sigma_{\tau}$
as \begin{equation}E_U=\frac{1}{2}UN_{\mathrm{orb}}\sum_{\tau}\left[(N_{\mathrm{orb}}-1)\mu_{\tau}^2-\sigma_{\tau}^2\right]\end{equation}and\begin{equation}E_{\mathrm{dc}}=\frac{1}{2}UN_{\mathrm{orb}}\sum_{\tau}\mu_{\tau}(N_{\mathrm{orb}}\mu_{\tau}-1).\end{equation}
A function of only $\mu_{\tau}$, $-E_{\mathrm{dc}}$ takes the same
form as $E_{\mathrm{fill}}$ except the prefactor on the quadratic term
is $N_{\mathrm{orb}}$ instead of 1; $E_U$ depends on both $\mu_{\tau}$
and $\sigma_{\tau}$.

In this work we are primarily interested in the formation energy,
which is defined
as \begin{equation}FE(x)=E(x)-[(1-x)E(0)+xE(1)],\label{fe}\end{equation}
where $E(x)$ is the cohesive energy of a system with intercalant
concentration $x$. The formation energy indicates whether a species of
intermediate $x$ has a higher or lower cohesive energy than the
corresponding linear combination of those of the endmembers.
Therefore, only energy terms beyond linear order in $x$ contribute.
The formation energy dictates the stability of such a species in the
limit of low temperature ($T\rightarrow0$); it is stable if negative
and unstable if positive. One can separately construct the formation
energy contributions stemming from $E_{\mathrm{DFT}}$,
$E_U-E_{\mathrm{dc}}$, $E_{\mathrm{fill}}$, $E_{\mathrm{ord}}$ using
expressions analogous expressions to Eq. \ref{fe}, allowing one to
understand the contribution of each term to the formation energy. Note
that $FE_{\mathrm{fill}}$ will be negative if $\mu(1-\mu)$ is
\emph{lower} than the endmember linear combination, while
$FE_{\mathrm{ord}}$ will be negative if $\sigma^2$ is \emph{higher}
than the endmember linear combination due to the negative sign in the
definition of $E_{\mathrm{ord}}$ in Eq. \ref{eord}. All formation
energies reported in this work are normalized per formula unit (f.u.).
We also compute the average intercalation voltage $V$
via \begin{equation} eV=E(\textrm{Li})+E(0)-E(1),\label{eq:voltage}
\end{equation} where $e$ is the elementary charge, $E(\textrm{Li})$ is the cohesive
energy of body-centered-cubic Li, and the energies are normalized to
the number of f.u.\cite{aydinol_ab_1997}

\section{Computational Details}\label{computational_details}

DFT+$U$ calculations based on the spin-dependent generalized gradient
approximation,\cite{perdew_generalized_1996} and the
rotationally-invariant Hubbard $U$
interaction\cite{liechtenstein_density-functional_1995} with $J$ set
to 0, and FLL dc are performed using the Vienna \textit{ab initio}
simulation package
(\textsc{vasp}).\cite{kresse_ab_1994,kresse_ab_1993,kresse_efficient_1996,kresse_efficiency_1996}
The projector augmented wave
method\cite{blochl_projector_1994,kresse_ultrasoft_1999} is employed
and the single-particle equations are solved with a plane-wave basis
set with a kinetic energy cutoff of 500 eV. We use a $9\times9\times9$
($6\times7\times8$) $k$-point grid for the rhombohedral (orthorhombic)
primitive unit cell of Li$_x$CoO$_2$ (Li$_x$FePO$_4$ and
Li$_x$CoPO$_4$) and $k$-point grids with approximately the same
$k$-point density for supercell calculations. The bulk lithium
calculation is performed using a $19\times19\times19$ $k$-point grid.
For structural relaxations in metals we employ the first-order
Methfessel-Paxton method\cite{methfessel_high-precision_1989} with a
50 meV smearing and for all other calculations the tetrahedron method
with Bl\"{o}chl corrections\cite{blochl_improved_1994} is used. The
ionic forces, stress tensor components, and total energy are converged
to 0.01 eV/\AA, 10$^{-3}$ GPa, and 10$^{-6}$ eV, respectively.

The disordered (solid solution) phase of Li$_{1/2}$CoO$_2$ is modeled
via the special quasirandom structure (SQS)
approach\cite{zunger_special_1990} using the Alloy Theoretic Automated
Toolkit
(\textsc{atat})\cite{van_de_walle_alloy_2002,van_de_walle_multicomponent_2009}
using point, pair, triplet, and quadruplet clusters. Candidate
structures are generated based on the correlation functions of
clusters with a maximum inter-site distance up to the in-plane 2nd
nearest neighbor and out-of-plane 1st nearest neighbor distance. To
evaluate the structures, we consider the minimal values of the
following figure of merit: the root-sum-square of the differences
between the cluster correlation functions and those of the random
structure. In this evaluation, the correlation functions of clusters
containing a maximum inter-site distance up to the in-plane 5th
nearest neighbor and inter-plane 6th nearest neighbor are taken into
account.

\section{Results and Discussion}

\subsection{Endmember electronic structure within DFT}\label{endmembers}

We begin by reviewing the electronic structure of the endmembers of
Li$_x$CoO$_2$ and Li$_x$FePO$_4$ within DFT, comparing with the latest
experimental understanding of these materials. LiCoO$_2$ crystallizes
in the layered structure illustrated in Fig. \ref{structures}(a)
consisting of layers of edge-sharing CoO$_6$ octahedra and layers of
Li. It has a rhombohedral primitive unit cell with the $R\bar{3}m$
space group and A-B-C (O3) oxygen
stacking.\cite{johnston_preparation_1958,orman_cobaltiii_1984} CoO$_2$
has a very similar structure with a hexagonal unit cell and A-B (O1)
oxygen stacking in the $P\bar{3}m1$ space
group.\cite{amatucci_coo2_1996} Here for convenience we consider
CoO$_2$ in the O3 structure, which has a very similar electronic
structure to that of the O1 structure within DFT (see the
Supplementary Material).

In Li$_x$CoO$_2$ the octahedral coordination of Co is slightly
distorted due to the ability of the oxygens to relax in the
out-of-plane direction, resulting in a symmetry lowering with $T_{2g}
\rightarrow A_{1g}+E_g'$. Nominally LiCoO$_2$ is in the $d^6$
configuration with filled $E_g'$ and $A_{1g}$ levels, while CoO$_2$
has one fewer electron ($d^5$).

\begin{figure*}[tb]
\begin{center}
\includegraphics[width=\textwidth]{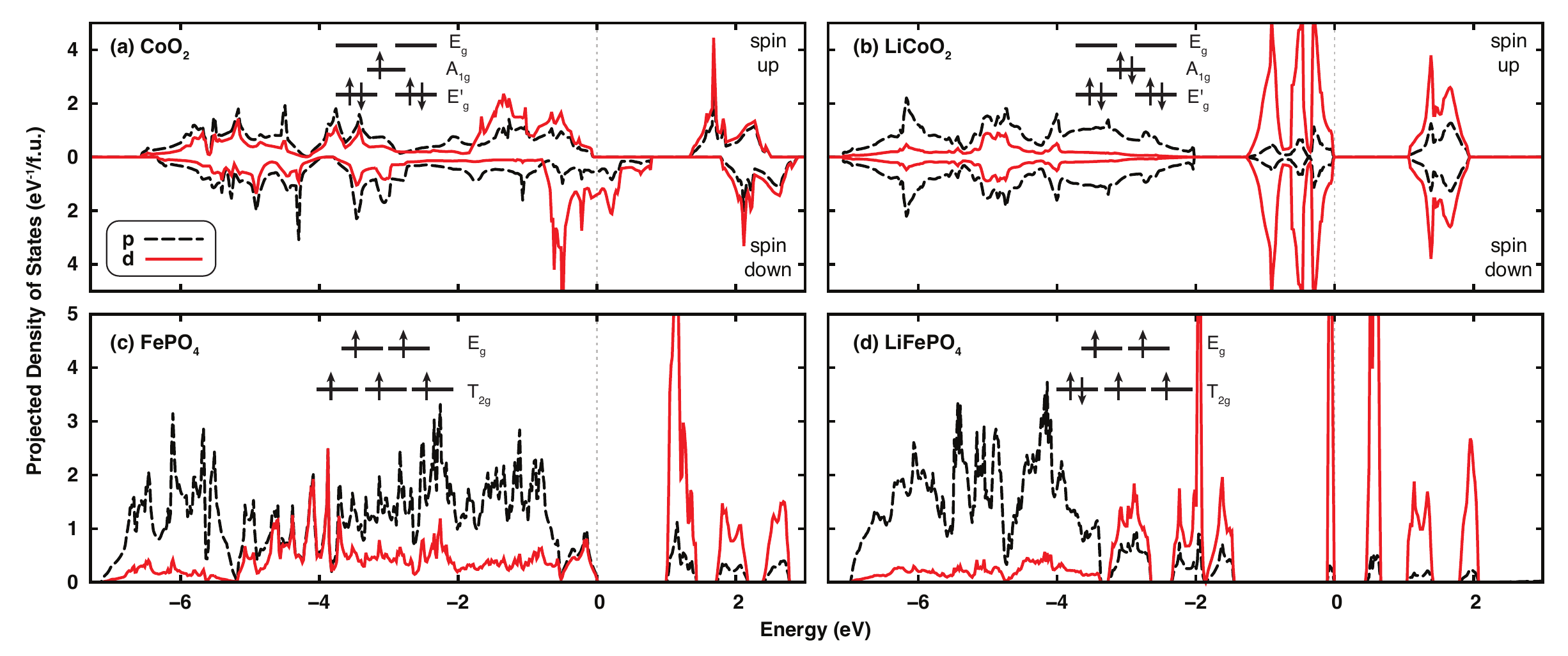}
\end{center}
\caption{Projected $p$ and $d$ density of states for (a) CoO$_2$, (b)
  LiCoO$_2$, (c), FePO$_4$, and (d) LiFePO$_4$ within DFT. For
  antiferromagnetic FePO$_4$ and LiFePO$_4$ only a single spin channel
  is shown. The dashed gray lines indicate the valence band maximum
  for insulators and the Fermi energy for metals. Insets are the
  nominal transition metal 3$d$ orbital fillings from crystal field
  theory.\label{pdos}}
\end{figure*}

The $p$ and $d$ projection of the electronic density of states within
DFT for CoO$_2$ and LiCoO$_2$ are shown in Fig. \ref{pdos}(a) and
\ref{pdos}(b), respectively. LiCoO$_2$ is found to be a band insulator
in agreement with
experiments.\cite{van_elp_electronic_1991,menetrier_insulator-metal_1999,menetrier_really_2008}
The computed band gap of 1.1 eV underestimates the experimental value
of 2.7 eV as is typical for DFT.\cite{van_elp_electronic_1991} The
occupancies of $A_{1g}$, $E_{g}'$ and $E_g$ are 0.94, 0.95, and 0.41
for LiCoO$_2$, demonstrating that hybridization between Co $d$ and O
$p$ states leads to appreciable occupation of the nominally-unoccupied
$E_g$ states. CoO$_2$ has a ferromagnetic low-spin metallic ground
state with a Co magnetic moment of 0.8 $\mu_B$. In experiments on
CoO$_2$, there is evidence for Fermi liquid behavior and Pauli
paramagnetism without any long-range magnetic
ordering.\cite{de_vaulx_electronic_2007,motohashi_synthesis_2007,kawasaki_measurement_2009}
For CoO$_2$ the occupation of $A_{1g}$, $E_{g}'$ and $E_g$ are 0.96
(0.67), 0.96 (0.76), and 0.54 (0.52) for spin up (down) electrons,
showing the significant degree of covalency in this system.

LiFePO$_4$ and FePO$_4$ both take on the olivine structure, which has
an orthorhombic primitive unit cell containing 4 formula units in the
$Pnma$ space
group.\cite{santoro_antiferromagnetism_1967,rousse_magnetic_2003} The
structure consists of corner-sharing FeO$_6$ octahedra layers
connected via PO$_4$ tetrahedra as shown in Fig. \ref{structures}(b).
For LiFePO$_4$ there are one-dimensional chains of Li ions.

Li$_x$FePO$_4$ has significantly distorted FeO$_6$ octahedra, though
for convenience we still crudely discuss the $d$ orbitals as $T_{2g}$
and $E_g$. Experimentally LiFePO$_4$ and FePO$_4$ are high-spin
antiferromagnetic (AFM) insulators with N\'{e}el temperatures of 52
and 125 K,
respectively.\cite{santoro_antiferromagnetism_1967,rousse_magnetic_2003}
While FePO$_4$ ($d^5$) nominally has all the $d$ orbitals on a given
site singly occupied with aligned spins (i.e., $S=5/2$), for
LiFePO$_4$ ($d^6$) there is one additional minority-spin electron in
the $T_{2g}$
manifold.\cite{tang_electronic_2003,shi_first-principles_2005} Fe
linked via corner-sharing octahedra in the same layer have
anti-aligned magnetic moments, while those laterally adjacent in
different layers linked via PO$_4$ have aligned magnetic
moments.\cite{santoro_antiferromagnetism_1967,rousse_magnetic_2003}

The projected density of states for the olivine endmembers are shown
in Fig. \ref{pdos}(c) and \ref{pdos}(d). Due to the
antiferromagnetism, both spin channels are identical so only one is
shown. FePO$_4$ can be viewed as a charge transfer type insulator
since the gap is $p$--$d$ in nature, whereas in LiFePO$_4$ $d$ states
form both the valence and conduction bands and the electronic
bandwidths near the Fermi energy are extremely narrow (as little as
0.1 eV). Although LiFePO$_4$ has an even number of electrons, the
local Coulomb interaction can play a strong role in developing or
enhancing the insulating behavior. Within DFT, the band gaps of
FePO$_4$ and LiFePO$_4$ are 1.0 eV and 0.4 eV, respectively. These
values are brought much closer to agreement with the experimental band
gaps of 1.9 eV\cite{zaghib_electronic_2007} and 3.8
eV\cite{zhou_electronic_2004,zaghib_electronic_2007} using the DFT+$U$
approach.\cite{zhou_electronic_2004,seo_calibrating_2015}

\subsection{Impact of $U$ on electronic structure of Li$_x$CoO$_2$}\label{lixcoo2_electronic_structure}

\begin{figure*}[tb]
\begin{center}
\includegraphics[width=\textwidth]{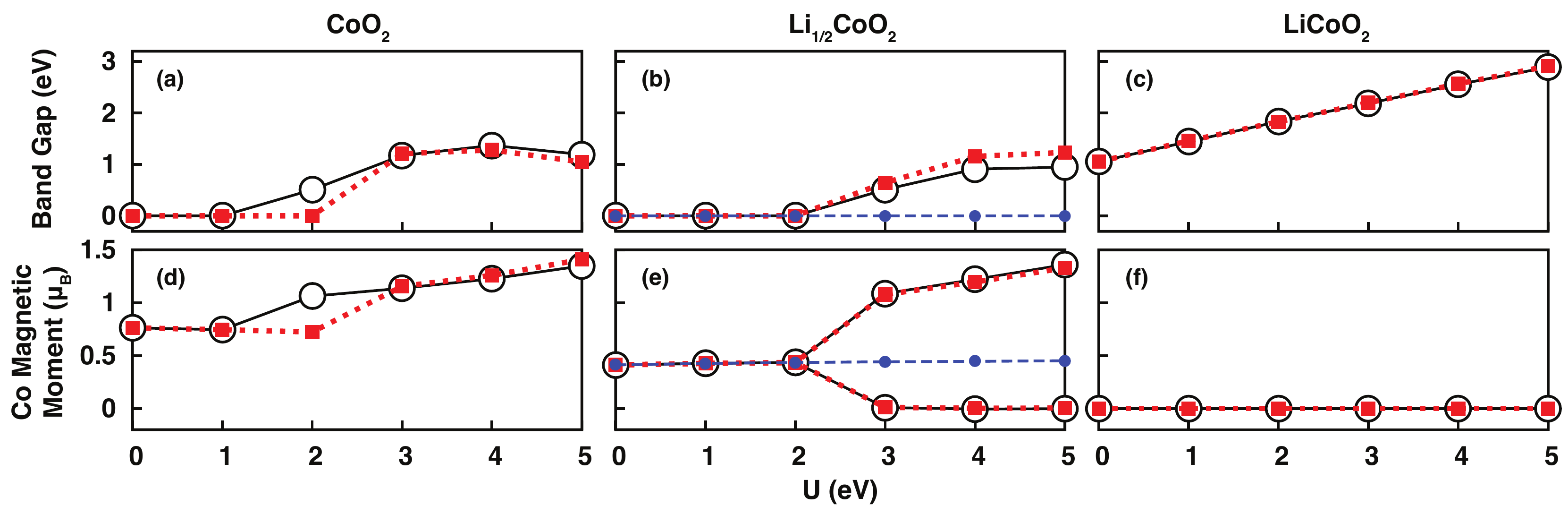}
\end{center}
\caption{Electronic band gap as a function of $U$ for (a) CoO$_2$, (b)
  Li$_{1/2}$CoO$_2$, and (c) LiCoO$_2$ with relaxations (filled red
  squares) and with frozen $U=0$ structures (large open black
  circles). The corresponding plots of Co magnetic moment are shown in
  panels (d)--(f). For panels (b) and (e), the additional small filled
  blue circles correspond to calculations with frozen $U=0$ structures
  and CO suppressed.\label{co_gaps}}
\end{figure*}

The on-site interaction $U$ has been computed as 4.9 and 5.4 eV for
LiCoO$_2$ and CoO$_2$, respectively.\cite{zhou_first-principles_2004}
The $U$-dependence of the band gap and Co magnetic moment for
Li$_x$CoO$_2$ are shown in Fig. \ref{co_gaps}. The band insulator
LiCoO$_2$ has no magnetic moment and its band gap increases roughly
linearly with $U$ from 1.1 eV at $U=0$ to 2.9 eV for $U=5$ eV.
Structural relaxations with $U$ have little impact on the electronic
structure. Note that we do not explore $U>5$ eV since in this regime
we find DFT+$U$ predicts a high-spin state for
LiCoO$_2$\cite{andriyevsky_electronic_2014} in contradiction with
experimental
observation\cite{menetrier_insulator-metal_1999,menetrier_really_2008}.

CoO$_2$ is semimetallic at lower values of $U$ with the $A_{1g}$ and
$E_g'$ states both partially occupied. Beyond $U=1$ eV (or $U=2$ eV
when including structural relaxations), an orbital ordering occurs in
which $E_g'$ completely fills and $A_{1g}$ becomes a nominally
half-filled $S=1/2$ state. This opens up a band gap of 0.5--1.2 eV and
increases the Co magnetic moment to 1.1--1.4 $\mu_B$, as compared to
0.7--0.8 $\mu_B$ for lower $U$.

The lowest-energy structure of Li$_{1/2}$CoO$_2$ has an in-plane
ordering of Li and vacancies corresponding to the unit cell shown in
Fig. \ref{structures}(a) with the Li1 ion
removed.\cite{reimers_electrochemical_1992,van_der_ven_first-principles_1998,wolverton_first-principles_1998}
Experimental studies suggest Li$_{1/2}$CoO$_2$ is a paramagnetic metal
with small Co magnetic moments of around 0.25--0.35
$\mu_B$.\cite{motohashi_electronic_2009,miyoshi_magnetic_2010,ou-yang_electronic_2012}
Within our calculations Li$_{1/2}$CoO$_2$ is a ferromagnetic metal for
$U\le2$ eV with equal Co magnetic moments of 0.4 $\mu_B$. For larger
$U$ values a new ground state with CO emerges in which the first site
takes on a CoO$_2$-like configuration with a moment of 1.1--1.4
$\mu_B$ and the second takes on a LiCoO$_2$-like configuration with no
moment (the actual charge difference between sites is small; see Fig.
\ref{site_avg}(a)). CO opens an electronic band gap that increases
with $U$ of 0.6--1.2 eV (0.5--0.9 eV without structural relaxations).
Ignoring structural relaxations, the metallic state without CO is
metastable and the Co magnetic moments are 0.4--0.5 $\mu_B$ and remain
roughly constant for all values of $U$ considered.

\subsection{Tendency for charge ordering in DFT+$U$}\label{charge_ordering}

To understand the tendency for CO in DFT+$U$, which is not observed in
experiment, we investigated whether its origin is the interaction
($U$) term or the dc term. Given that the dc energy has a term which
is a negative quadratic of $N_d$, it has a disposition towards charge
ordering. Alternatively, the interaction term may indirectly drive CO
as a means to reduce the interaction penalty, arising at the expense
of $E_{DFT}$. To clearly identify the origin of CO, we implemented a
modified DFT+$U$ approach in which both the single-particle potential
and total energy contributions stemming from the interaction term or
the dc term are averaged over correlated sites. We call this
site-averaged interaction and site-averaged dc, respectively.

\begin{figure}[htbp]
\begin{center}
\includegraphics[width=\linewidth]{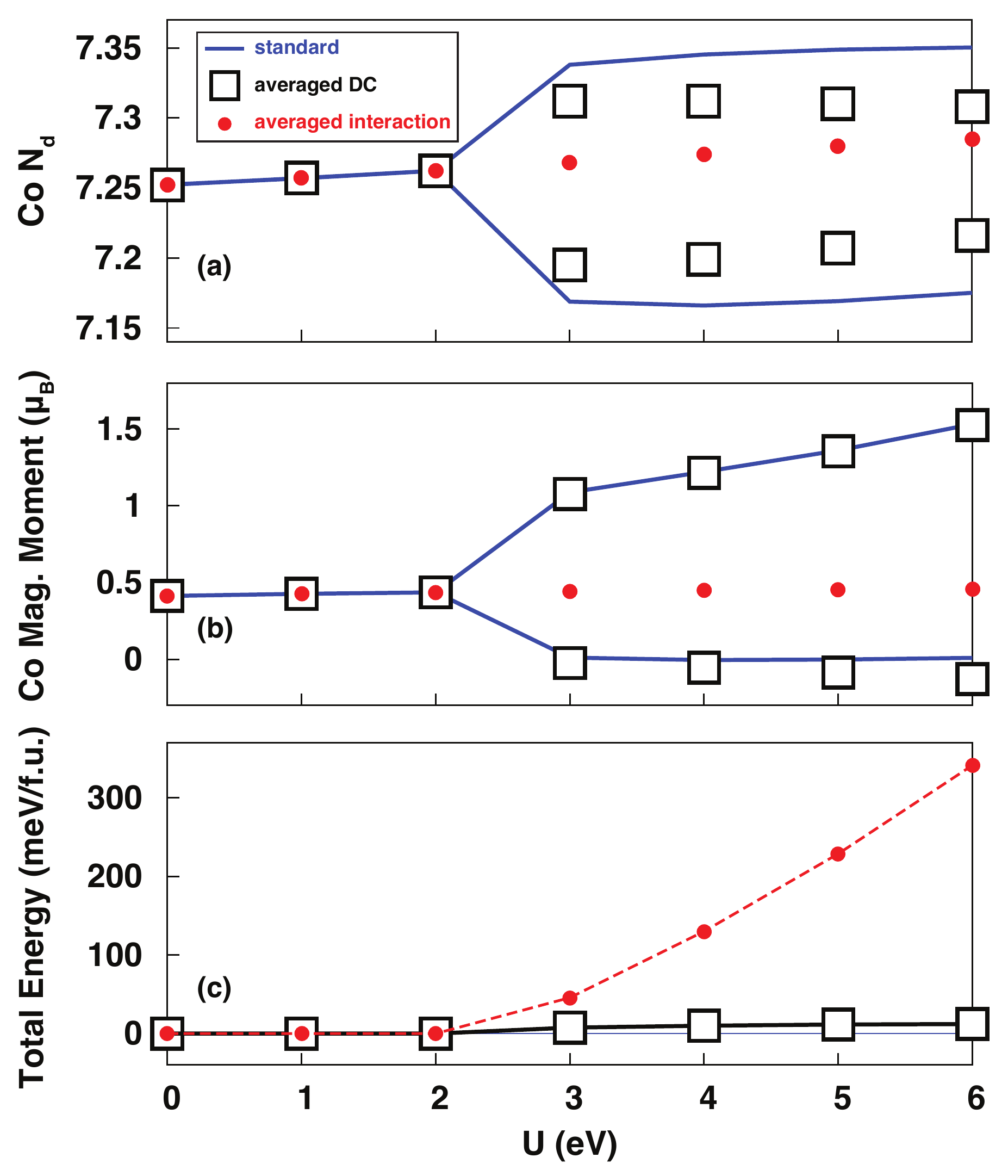}
\end{center}
\caption{(a) Co N$_d$, (b) Co magnetic moments, and (c) total energy
  for Li$_{1/2}$CoO$_2$ with the frozen $U=$ 0 eV structure for
  standard DFT+$U$ (blue lines) as well as with the interaction
  (filled red circles) or double counting (open black squares) terms
  averaged over correlated sites. Total energies in panel (c) are with
  respect to those of the standard DFT+$U$.\label{site_avg}}
\end{figure}

Figure \ref{site_avg} illustrates the results of this computational
experiment for ordered Li$_{1/2}$CoO$_2$ using the frozen $U=0$
structure. With site-averaged dc (black squares) we still find a CO
transition, for $U>2$ eV. The magnitude of the CO in terms of $N_d$ is
slightly reduced, but the deviations in Co magnetic moment are the
same or even more substantial than the standard DFT+$U$ results (blue
lines). The total energy shown in panel (c) for this case is only
slightly (on the order of 10 meV/f.u.) higher than in the case of
standard DFT+$U$, which indicates that the dc energetics are not very
much lowered via CO. With site-averaged interaction (red circles),
however, no CO can be obtained and the total energy is massively
penalized as $U$ increases (by hundreds of meV/f.u.). Therefore, we
conclude that it is the interaction term and not the dc term that is
responsible for the CO in DFT+$U$. This suggests that more accurate
solutions to the interaction problem such as DMFT may be critical to
resolving the issues associated with spurious CO.

\subsection{Impact of $U$ on phase stability of Li$_x$CoO$_2$}\label{lixcoo2_phase_stability}

\begin{figure}[htbp]
\begin{center}
\includegraphics[width=\linewidth]{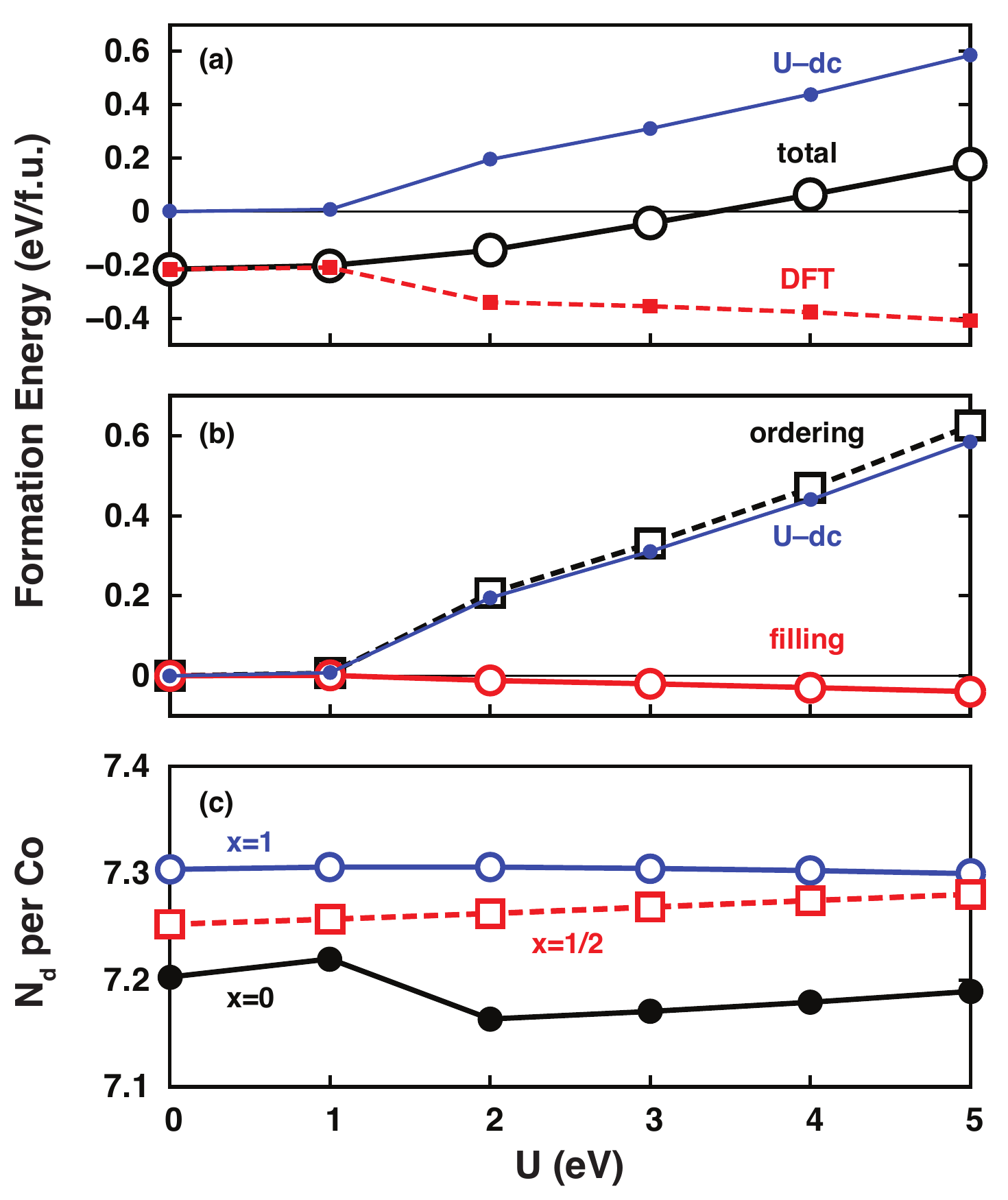}
\end{center}
\caption{(a) Total Li$_{1/2}$CoO$_2$ formation energy (open black
  circles) and its DFT (filled red squares) and $U$--dc (filled blue
  circles) components as a function of $U$ (b) Li$_{1/2}$CoO$_2$
  $U$--dc (filled blue circles) formation energy component and its
  orbital filling (open red circles) and orbital ordering (open black
  squares) components as a function of $U$ (c) Number of $d$ electrons
  per Co as a function of $U$ for CoO$_2$ (filled black circles),
  Li$_{1/2}$CoO$_2$ (open red squares), and LiCoO$_2$ (open blue
  circles). All data correspond to the case of frozen $U=0$ structures
  and CO suppressed in Li$_{1/2}$CoO$_2$.\label{co_frozen_decomp}}
\end{figure}

We first consider the formation energy of Li$_{1/2}$CoO$_2$ in the
frozen $U=0$ structure (i.e., no structural relaxations when imposing
$U$) and without allowing CO, in order to purely see the effects of
$U$ in the absence of CO and lattice distortions. For the experimental
Li ordering in Li$_{1/2}$CoO$_2$, the two Co atoms in the unit cell
are equivalent by point symmetry. Therefore CO is a spontaneously
broken symmetry, which enables one to precisely investigate various
observables with and without CO. We will return to the effects of both
CO and structural relaxations after thoroughly explaining the role of
$U$ in their absence.

As shown in Fig. \ref{co_frozen_decomp}(a), Li$_{1/2}$CoO$_2$ is phase
stable with a total formation energy of $-217$ meV for $U=0$. The
formation energy increases monotonically with $U$ and for $U>3$ eV it
becomes positive, corresponding to a prediction of phase separation.
This indicates that the trend of a destabilization of compounds of
intermediate $x$ found previously in
Li$_x$FePO$_4$\cite{zhou_phase_2004} also occurs for Li$_x$CoO$_2$.
Furthermore, it demonstrates that such a trend is found even in the
absence of CO.

To illustrate the origin of this behavior, we also examine the DFT and
$U$--dc components of the total formation energy in Fig.
\ref{co_frozen_decomp}(a). For $U=1$ eV the $U$--dc component is
negligible ($<0.01$ eV) and the slightly less negative value of
formation energy ($-203$ meV) stems from the DFT component. Compared
to the $U=0$ values, for $U=1$ eV the DFT energy increases by 17 meV
for Li$_1/2$CoO$_2$ compared to only 7 meV for CoO$_2$ and 14 meV for
LiCoO$_2$. For larger $U$ the $U$--dc component is strongly positive
and increases roughly linearly with $U$ at a rate of around 130 meV
per eV, leading to a more rapid increase in total formation energy.
The DFT component has the opposite trend of becoming more negative
with $U$, largely since the DFT energy of CoO$_2$ is strongly
penalized by the orbital ordering, but the overall effect on the
formation energy is smaller with changes of around 15--30 meV per eV.
Therefore, it is the $U$--dc component that is responsible for the
destabilization of Li$_{1/2}$CoO$_2$.

The number of $d$ electrons per Co site ($N_d$) is plotted for $x=0$,
$x=1/2$, and $x=1$ as a function of $U$ in Fig.
\ref{co_frozen_decomp}(c). We note that the difference in $N_d$
between CoO$_2$ and LiCoO$_2$ is only around 0.1 even though Li is
nominally donating a full electron, which is due to the $p$--$d$
rehybridization effect in
Li$_x$CoO$_2$.\cite{wolverton_first-principles_1998,marianetti_role_2004}
While the behavior of $N_d$ is roughly constant at 7.3 for LiCoO$_2$
and smoothly increasing for Li$_{1/2}$CoO$_2$ between 7.25 and 7.28,
there is discontinuous behavior for CoO$_2$ in which $N_d$ drops from
7.22 to 7.16 from $U=1$ to $U=2$ eV corresponding to the orbital
ordering. This change in electronic structure in the $x=0$ endmember
is responsible for the change in behavior in the $U$--dc formation
energy contribution.

Since the dc term in Eq. \ref{flldc} is a negative quadratic function,
one might expect that the dc is responsible for the trend towards
phase separation. This simple line of reasoning immediately becomes
more complicated given that $N_d$ is a nonlinear function of $x$ as
demonstrated in Fig. \ref{co_frozen_decomp}(c), and a careful analysis
in Sec. \ref{strawman} shows that it is not the dc that drives phase
separation. Alternatively, we proceed to understand the contributions
of both the $E_U$ and $E_{\mathrm{dc}}$ terms simultaneously in a
different framework using the energy decomposition described in Sec.
\ref{decompositioneqns}. In Fig. \ref{co_frozen_decomp}(b) we break
down the $U$--dc formation energy contribution into the filling and
ordering contributions. Remarkably, the magnitude of the filling
contribution contributes negligibly, only being at most tens of meV in
magnitude, whereas essentially all of the $U$--dc formation energy
comes from the ordering term. Therefore, it is the ordering rather
than the filling of the correlated $d$ orbitals that drives phase
separation.

\begin{figure}[htbp]
\begin{center}
\includegraphics[width=\linewidth]{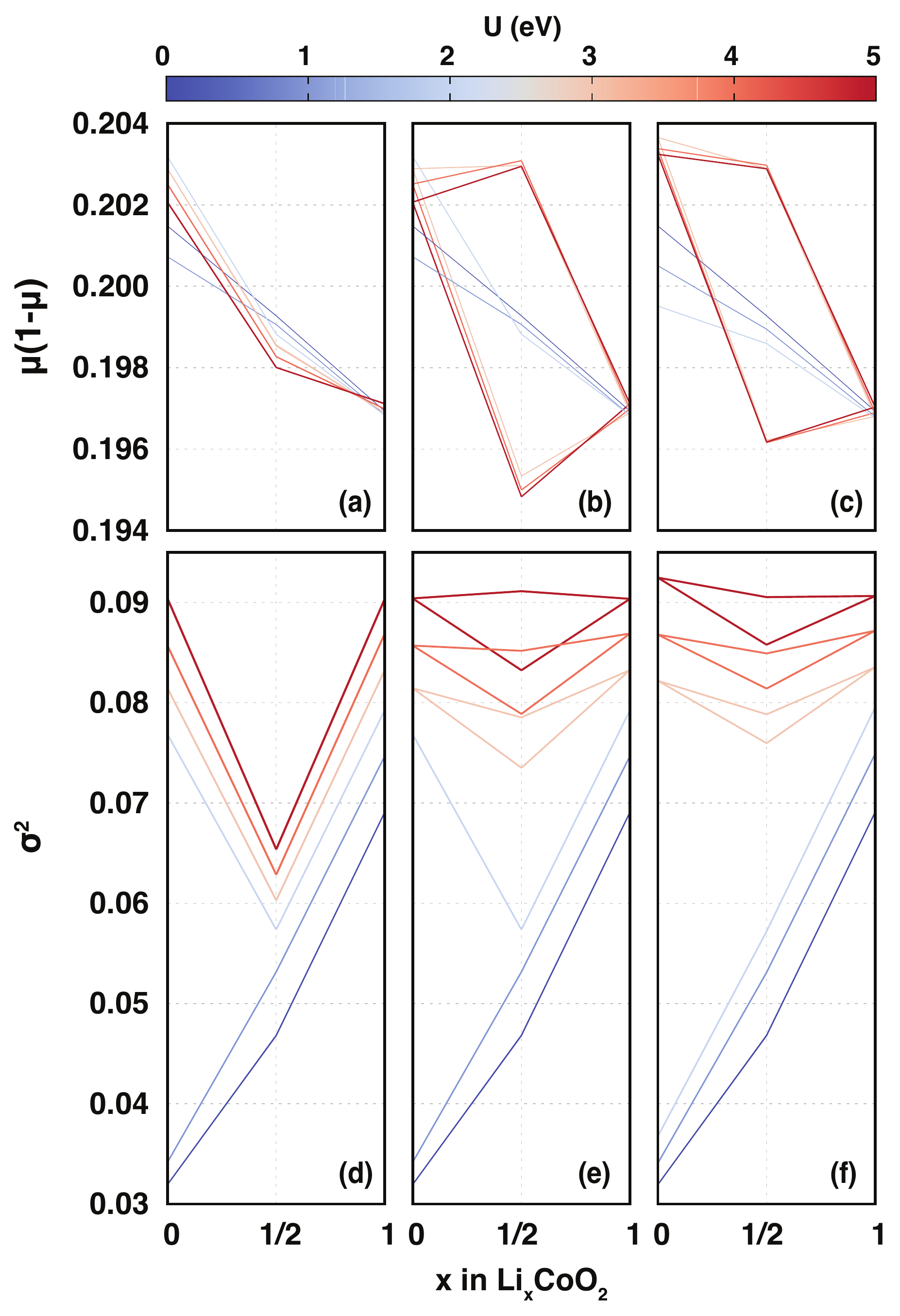}
\end{center}
\caption{Filling factor $\mu(1-\mu)$ as a function of $x$ for
  different $U$ in Li$_x$CoO$_2$ for (a) frozen $U=0$ structures and
  CO suppressed, (b) frozen $U=0$ structures and CO allowed, and (c)
  relaxed structures. Panels (d)--(f) show the corresponding plots for
  the ordering factor $\sigma^2$. The two lines per $U$ in some plots
  correspond to the two distinct Co sites in Li$_{1/2}$CoO$_2$. The
  line thickness increases for increasing values of
  $U$.\label{co_mu_sigma}}
\end{figure}

The individual filling factor $\mu(1-\mu)$ and ordering factor
$\sigma^2$ as a function of Li concentration for different $U$ are
shown for this case in Fig. \ref{co_mu_sigma}(a) and
\ref{co_mu_sigma}(d), respectively. We note that based on nominal
electron counting $\mu(1-\mu)$ will be 0.25 for Co$^{4+}$ and 0.24 for
Co$^{3+}$. Consistent with this expectation, we observe that the
filling factor is highest for $x=0$ and lowest for $x=1$. The actual
values for Li$_x$CoO$_2$ are lower in magnitude (around 0.2) due to
the substantial covalent nature of the bonding, in particular the
hybridization of O $p$ states with Co $E_g$ states. Although this
filling factor magnitude is around half an order of magnitude higher
than that of the ordering factor, the relative deviations of the
$x=1/2$ value compared to the average of the $x=0$ and $x=1$ values
are tiny (around $10^{-3}$ electrons). This is why filling term leads
to a negligible contribution to the formation energy.

The values of $\sigma^2$ increase with $U$ for all $x$ in agreement
with the expectation that the $U$ and dc parts of the total energy
functional will penalize fractional orbital occupancy (i.e.,
$0<n_m^{\tau s}<1$). For $U\le1$ eV $\sigma^2$ is nearly linear in
$x$, thus leading to no appreciable formation energy contribution.
However, once CoO$_2$ undergoes the orbital ordering, its value of
$\sigma^2$ significantly increases from 0.03 to 0.08. After this phase
transition, $\sigma$ for Li$_{1/2}$CoO$_2$ is substantially lower than
the average of those of CoO$_2$ and LiCoO$_2$. For example, for $U=2$
eV $\sigma$ is 0.277 for CoO$_2$ and 0.282 for LiCoO$_2$ but only
0.239 for Li$_{1/2}$CoO$_2$. This lower-than-average $\sigma^2$ is
translated to a positive formation energy contribution via the
negative sign in prefactor of Eq. \ref{eord}.

\begin{figure}[htbp]
\begin{center}
\includegraphics[width=\linewidth]{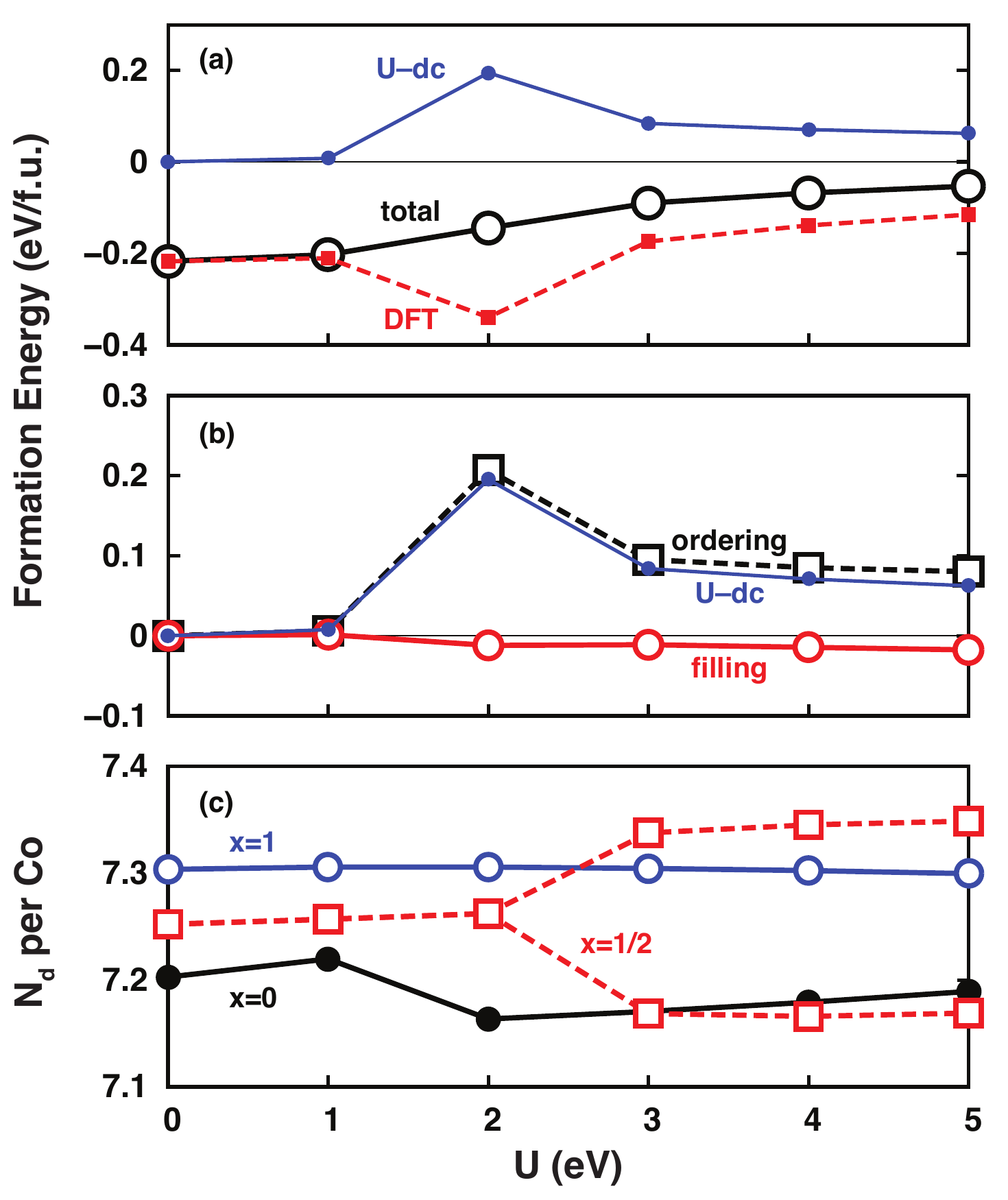}
\end{center}
\caption{(a) Total Li$_{1/2}$CoO$_2$ formation energy (open black
  circles) and its DFT (filled red squares) and $U$--dc (filled blue
  circles) components as a function of $U$ (b) Li$_{1/2}$CoO$_2$
  $U$--dc (filled blue circles) formation energy component and its
  orbital filling (open red circles) and orbital ordering (open black
  squares) components as a function of $U$ (c) Number of $d$ electrons
  per Co as a function of $U$ for CoO$_2$ (filled black circles),
  Li$_{1/2}$CoO$_2$ (open red squares), and LiCoO$_2$ (open blue
  circles). All data correspond to the case of frozen $U=0$ structures
  and CO allowed and the two lines in panel (c) for Li$_{1/2}$CoO$_2$
  correspond to the two distinct Co sites.\label{co_frozen_co_decomp}}
\end{figure}

The fundamental behavior we find is that $U$ drives phase separation
via enhanced ordering of the correlated orbitals for the endmembers
relative to the species with intermediate Li concentration. Given that
we restricted the possibility of CO, Li$_{1/2}$CoO$_2$ has no low
energy means to orbitally order and keep pace with CoO$_2$ as $U$ is
increased. We will articulate this important point in a bit more
detail. For $U\le1$ eV $\sigma^2$ is much smaller for $x=0$ than
$x=1$; CoO$_2$ has a smaller range of $n_m^{\tau s}$ due to its
semimetallic behavior and enhanced hybridization with O $p$ states.
Once it has orbitally ordered at $U=2$ eV and the hole in the $T_{2g}$
manifold is localized in the minority-spin $A_{1g}$ state, however,
CoO$_2$ has 5 very occupied $d$ orbitals ($n_m^{\tau s}\ge0.96$) and 5
much less occupied orbitals ($n_m^{\tau s}$ of around 0.15 for
minority-spin $A_{1g}$ and higher values of 0.49--0.59 for $E_g$ due
to $p$--$d$ hybridization). LiCoO$_2$ is a band insulator, so there is
no abrupt change in $\sigma^2$ as a function of $U$. In terms of
$\sigma^2$, LiCoO$_2$ behaves very similarly to CoO$_2$ in the regime
of $U$ in which CoO$_2$ is orbitally ordered. In LiCoO$_2$ there are 6
very occupied $d$ orbitals ($n_m^{\tau s}\approx0.96$ for $A_{1g}$ and
$E_g'$ states) and 4 much less occupied orbitals ($n_m^{\tau s}=0.39$
for $E_g$ states). This gives a large spread ($\sigma$) in orbital
occupancies for LiCoO$_2$ in addition to CoO$_2$. In contrast, for
Li$_{1/2}$CoO$_2$ we have a metallic state with a nominally
half-filled minority-spin $A_{1g}$ level ($n_m^{\tau s}\approx0.61$)
and thus smaller $\sigma^2$. This lower-than-average $\sigma^2$ for
Li$_{1/2}$CoO$_2$ is what results in a positive contribution to the
formation energy from the $U$--dc energetics.

This same type of behavior is preserved even when we now allow CO in
Li$_{1/2}$CoO$_2$, though CO will allow Li$_{1/2}$CoO$_2$ to order and
completely avoid phase separation for all $U$ (we still restrict the
possibility of structural relaxations until later in this analysis).
Fig. \ref{co_frozen_co_decomp} shows the results with the frozen $U=0$
structures but now allowing for CO in Li$_{1/2}$CoO$_2$. Here again
the formation energy increases with $U$. For small values of $U$ the
increase is small and stems from the DFT contribution. After CoO$_2$
orbitally orders and opens a band gap, the $U$--dc energetics are a
phase separating contribution to the total formation energy. As
before, essentially all of the $U$--dc contribution comes from the
ordering, not the filling, of the $d$ orbitals. When Li$_{1/2}$CoO$_2$
charge orders for $U$ greater than 2 eV, the $U$--dc phase separating
contribution is significantly dampened but there is always still a
positive phase separating contribution (53--84 meV). CO also leads to
an increase in the DFT formation energy contribution from $-0.34$ eV
at $U=2$ eV to $-0.17$ eV at $U=3$ eV.

As illustrated in Fig. \ref{co_mu_sigma}(b) and \ref{co_mu_sigma}(e),
the variations in filling factor $\mu(1-\mu)$ are again negligible so
the only appreciable component to the $U$--dc energetics stems from
the changes in ordering factor $\sigma^2$. The CO of Li$_{1/2}$CoO$_2$
significantly increases the average $\sigma^2$, but it still always
lags behind the average of those of CoO$_2$ and LiCoO$_2$. For
example, for $U=5$ eV $E_{\mathrm{ord}}$ is 87 meV for $x=1/2$ and 90
meV for the average for the $x=0$ and $x=1$ values. In this case we
still end up with a total formation energy that steadily increases
with $U$, though now CO dampens the process substantially such that
for $U=5$ eV the value remains negative consistent with experiment.

\begin{figure}[htbp]
\begin{center}
\includegraphics[width=\linewidth]{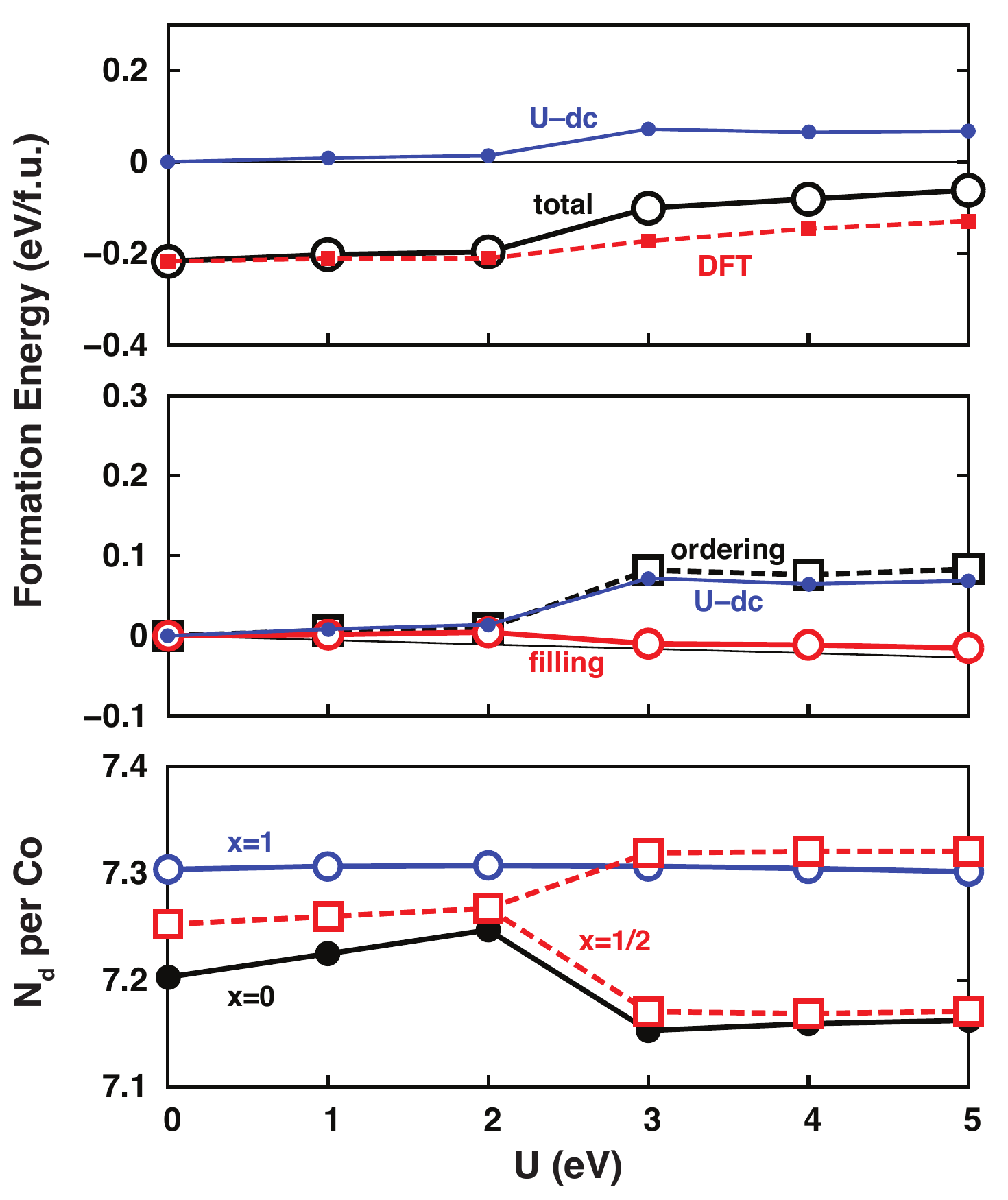}
\end{center}
\caption{(a) Total Li$_{1/2}$CoO$_2$ formation energy (open black
  circles) and its DFT (filled red squares) and $U$--dc (filled blue
  circles) components as a function of $U$ (b) Li$_{1/2}$CoO$_2$
  $U$--dc (filled blue circles) formation energy component and its
  orbital filling (open red circles) and orbital ordering (open black
  squares) components as a function of $U$ (c) Number of $d$ electrons
  per Co as a function of $U$ for CoO$_2$ (filled black circles),
  Li$_{1/2}$CoO$_2$ (open red squares), and LiCoO$_2$ (open blue
  circles). All data correspond to the case of fully relaxed
  structures and the two lines in panel (c) for Li$_{1/2}$CoO$_2$
  correspond to the two distinct Co sites.\label{co_rel_decomp}}
\end{figure}

When we include CO in Li$_{1/2}$CoO$_2$ and full structural
relaxations, we find the same fundamental effect as in the previous
case only allowing CO: structural relaxations only provide a further
dampening. As shown in Fig. \ref{co_rel_decomp}, the total formation
energy increases with $U$ and the positive contribution stems from the
ordering component of the $U$--dc energetics. Here the CO and CoO$_2$
orbital ordering both occur for $U>2$ eV. The increase in the DFT
formation energy contribution upon CO is dampened due to relaxations.
The total formation energy remains negative at $-61$ meV at $U=5$ eV.
The plots of $\mu(1-\mu)$ and $\sigma^2$ are shown in Fig.
\ref{co_mu_sigma}(c) and \ref{co_mu_sigma}(f), respectively.
Structural relaxations serve to slightly enhance $\sigma^2$ for
CoO$_2$ and decrease the difference in $\sigma^2$ for the distinct Co
sites in Li$_{1/2}$CoO$_2$.

These results have significant implications on the accuracy and
robustness of the DFT+$U$ description of strongly correlated
materials. Without CO, which is not found in experiments on this class
of materials, DFT+$U$ incorrectly predicts that Li$_x$CoO$_2$ phase
separates once $U$ becomes appreciable. This is true even if we allow
structural relaxations while suppressing CO, which is possible since
CO is a spontaneously broken symmetry in Li$_{1/2}$CoO$_2$, and in
this case the total formation energy values are -40 meV, +67 meV, and
+182 meV for $U$ of 3, 4, and 5 eV, respectively. That DFT+$U$
requires artificial CO to correctly capture the phase stable nature of
Li$_x$CoO$_2$ is a significant weakness of this approach.

\subsection{Impact of $U$ on electronic structure of Li$_x$FePO$_4$}\label{lixfepo4_electronic_structure}

\begin{figure*}[tb]
\begin{center}
\includegraphics[width=\textwidth]{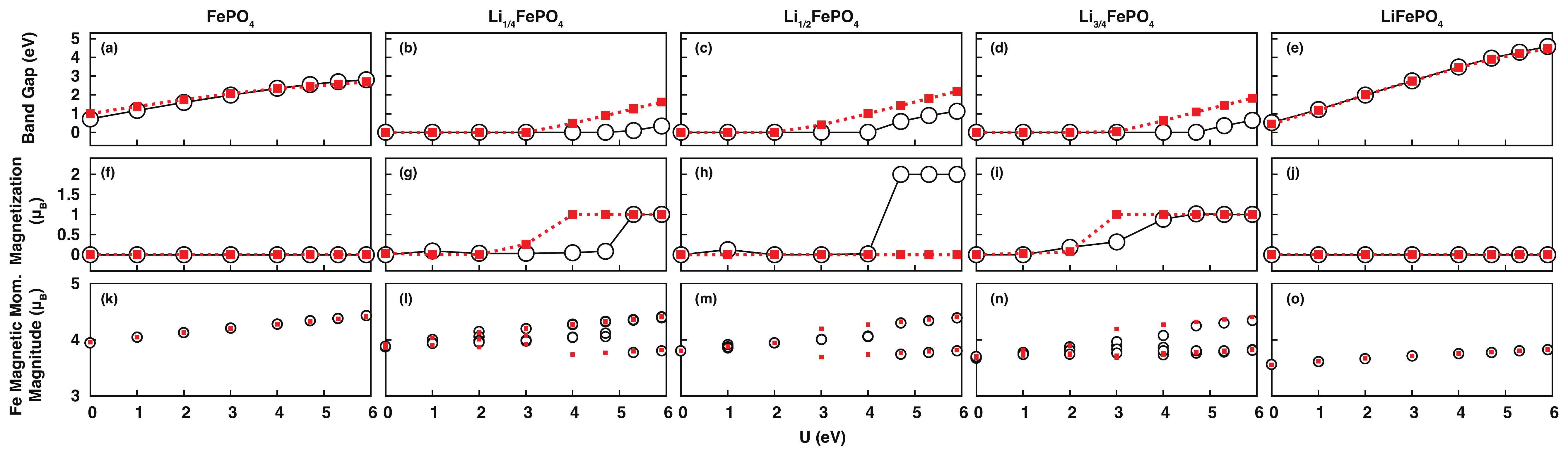}
\end{center}
\caption{Electronic band gap as a function of $U$ for (a) FePO$_4$,
  (b) Li$_{1/4}$FePO$_4$, (c) Li$_{1/2}$FePO$_4$, (d)
  Li$_{3/4}$FePO$_4$, and (e) LiFePO$_4$ with relaxations (filled red
  squares) and with frozen linearly interpolated experimental
  structures (open black circles). The corresponding plots of total
  magnetization and Fe magnetic moment magnitudes are shown in panels
  (f)--(j) and (k)--(o), respectively. For panels (k)--(o), the
  multiple symbols per $U$ correspond to the 4 distinct Fe
  sites.\label{fe_gaps}}
\end{figure*}

The interaction $U$ has been computed as 3.7 and 4.9 eV for LiFePO$_4$
and FePO$_4$, respectively.\cite{zhou_first-principles_2004} The
variation of the band gap, total magnetization, and Fe magnetic moment
as a function of $U$ for Li$_x$FePO$_4$ is illustrated in Fig.
\ref{fe_gaps}. The endmembers are both AFM so there is zero net
magnetization. The Fe magnetic moment, which increases approximately
linearly with $U$, is 4.0--4.4 $\mu_B$ for FePO$_4$ and 3.6--3.8
$\mu_B$ for LiFePO$_4$ for 0 eV $\le U\le$ 6 eV. The endmembers are
both insulating. For this $U$ range the band gaps increases from 1.0
to 2.7 eV for FePO$_4$ and 0.4 to 4.5 eV, a much larger range, for
LiFePO$_4$. Very similar results for the endmembers are found when one
freezes the ions to the experimental structures.

For intermediate $x$ we consider the lowest-energy configurations that
fit within the primitive unit cell shown in Fig. \ref{structures}(b).
The structures correspond to removing Li2, Li3, and Li4 for $x=1/4$,
Li3 and Li4 for $x=1/2$, and Li4 for $x=3/4$.\cite{zhou_phase_2004}
Calculations on the two other possible primitive cell
Li$_{1/2}$FePO$_4$ structures are always found to be higher in energy,
so we do not discuss them here. In addition to the case in which
structures are fully relaxed, we also perform calculations on
structures constructed via linear interpolation of the experimental
$x=0$ and $x=1$ structures. This is an ideal manner to isolate the
effect of lattice relaxations given that appreciable lattice
distortions arise in the intermediate compounds even for $U=0$.

For $U=0$ the intermediate $x$ species of Li$_x$FePO$_4$ are AFM
metals without any CO. The Fe magnetic moments are 3.9, 3.8, and 3.7
$\mu_B$ for $x=1/4$, $x=1/2$, and $x=3/4$, respectively. The magnetic
moment magnitude gradually increases with $U$, and above critical $U$
values a CO transition occurs leading to distinct Fe$^{3+}$-like
($d^5$) and Fe$^{2+}$-like ($d^6$) sites within the primitive unit
cell. This symmetry breaking leads to distinct local magnetic moments
and the opening of an electronic band gap. Unlike Li$_x$CoO$_2$ for
which $N_d$ differences among correlated sites are 0.15-0.18 electrons
in the CO state, Li$_x$FePO$_4$ has CO states with substantially
larger $N_d$ differences around 0.4 electrons due to the highly
localized nature of this system.

For Li$_{1/2}$FePO$_4$ using the frozen structure (linear
interpolation of endmember experimental structures), the CO transition
at $U>4$ eV yields a ferrimagnetic state with total magnetization 2
$\mu_B$; there are 2 Fe magnetic moments of 3.7 $\mu_B$
(Fe$^{2+}$-like) and 2 Fe magnetic moments of $-4.3$ $\mu_B$
(Fe$^{3+}$-like). It opens a band gap of 0.6 eV for $U=4.7$ eV that is
further increased with $U$ to a value of 1.1 eV for $U=5.9$ eV.
Including structural relaxations has little effect on the electronic
structure in the regime of $U$ before the onset of CO. However,
relaxations assist the CO transition and result in a lower critical
value of $U=2$ eV above which the CO state remains AFM with Fe
magnetic moments of $\pm$ 3.7 and $\pm$ 4.2 $\mu_B$. In addition,
relaxation serves to enhance the band gap of the CO state to values of
0.4--2.2 eV.

Li$_{1/4}$FePO$_4$ and Li$_{3/4}$FePO$_4$ show similar behavior in
which at critical values of $U$ (lower when structural relaxation is
included) a CO transition opens an electronic band gap. For
Li$_{1/4}$FePO$_4$ with (without) relaxations a band gap opens at
$U=4$ eV (5.3 eV) when CO yields 1 Fe$^{3+}$-like site and 3
Fe$^{2+}$-like sites. For Li$_{3/4}$FePO$_4$ with (without)
relaxations a band gap opens at $U=3$ eV (5.3 eV) when CO yields 3
Fe$^{3+}$-like sites and 1 Fe$^{2+}$-like site. For $x=1/4$ and
$x=3/4$ the magnitude of the band gap is around 0.1--0.6 eV for the
frozen structures and a larger values of 0.5--1.8 eV including
relaxations. For these $x$, the CO transition always leads to a
ferrimagnetic state with magnetization of 1 $\mu_B$. The Fe magnetic
moments of the distinct sites are around 3.8 and 4.4 $\mu_B$ and
slowly increase with $U$ as in the case of $x=1/2$. Differences in the
magnetic moments between the frozen and relaxed structures in the CO
regime are small. We note that for these $x$, unlike in $x=1/2$, we
find partial CO for intermediate values of $U$ in which the Fe
magnetic moments begin to take on slightly different values without
the presence of a band gap. For example, including relaxations for
$U=1$ eV Li$_{1/4}$FePO$_4$ is metallic with Fe magnetic moments of
3.9, $-3.9$, $-4.0$, and 4.0 $\mu_B$. This indicates DFT+$U$ is
driving Li$_x$FePO$_4$ towards CO even for small $U$, which is to be
expected given the very narrow bandwidths.

Unlike for Li$_x$CoO$_2$ (see Fig. \ref{co_gaps}), for Li$_x$FePO$_4$
the CO is stabilized by structural relaxations as evidenced by the
lower critical $U$ values for CO. The stabilization of CO by
structural relaxations suggests stronger coupling between the
electronic and lattice degrees of freedom in Li$_x$FePO$_4$ as
compared to Li$_x$CoO$_2$. This is consistent with the evidence for
polarons in
Li$_x$FePO$_4$.\cite{ellis_small_2006,zaghib_electronic_2007}


\subsection{Impact of $U$ on phase stability of Li$_x$FePO$_4$}\label{lixfepo4_phase_stability}

\begin{figure*}[tb]
\begin{center}
\includegraphics[width=\textwidth]{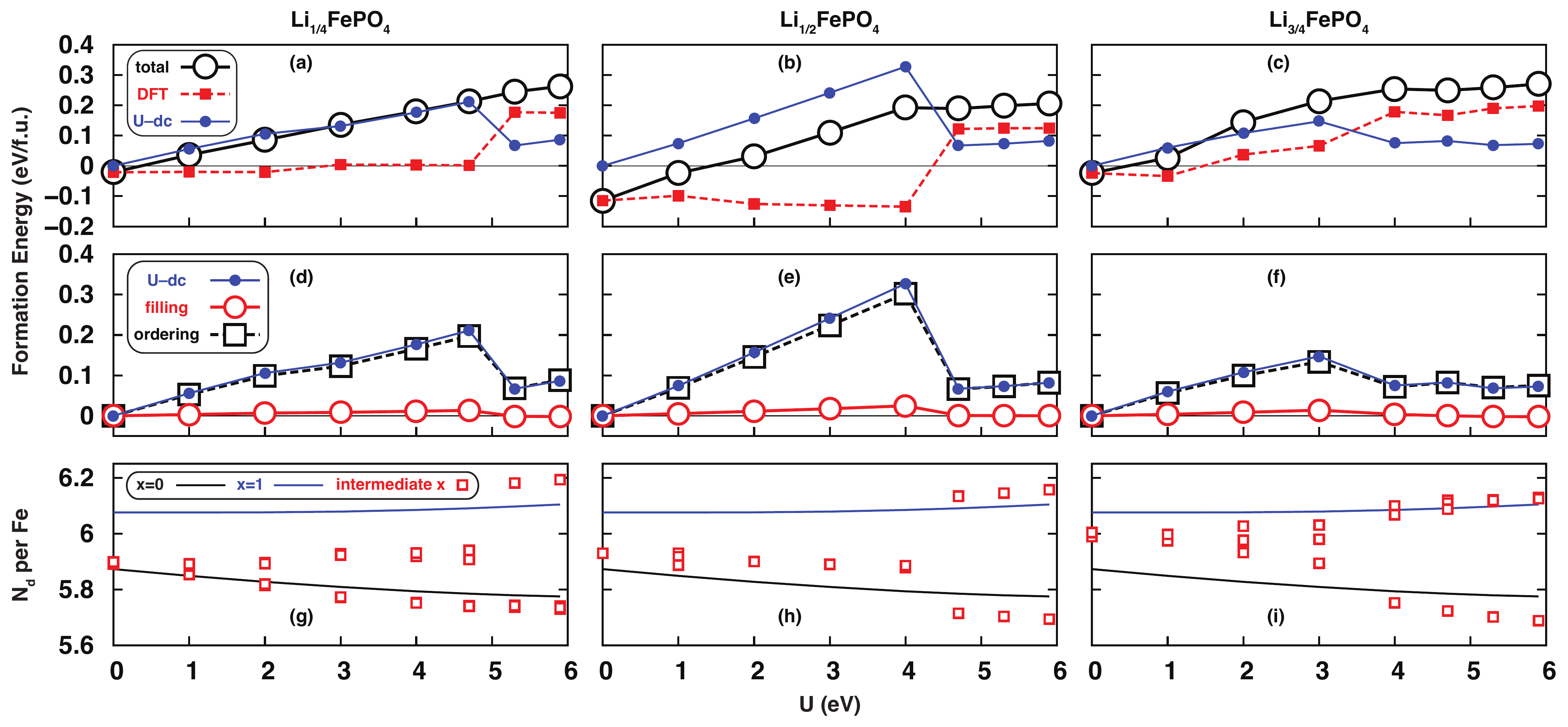}
\end{center}
\caption{Total Li$_x$FePO$_4$ formation energy (open black circles)
  and its DFT (filled red squares) and $U$--dc (filled blue circles)
  components as a function of $U$ for (a) $x=1/4$, (b), $1/2$, and (c)
  $x=3/4$. (d)--(f) show the corresponding plots of Li$_x$FePO$_4$
  $U$--dc (filled blue circles) formation energy component and its
  orbital filling (open red circles) and orbital ordering (open black
  squares) components as a function of $U$. (g)--(i) show the
  corresponding plots of number of $d$ electrons per Fe as a function
  of $U$ for FePO$_4$ (black line), Li$_x$FePO$_4$ (open red squares),
  and LiFePO$_4$ (blue line). All data correspond to the case of
  frozen linearly interpolated experimental
  structures.\label{fe_interp_decomps}}
\end{figure*}

The formation energy behavior for Li$_x$FePO$_4$ as a function of $U$
without allowing for the effect of structural relaxations is
summarized in Fig. \ref{fe_interp_decomps}. The behavior shares many
similarities and a few differences to that of Li$_x$CoO$_2$, which we
will describe. As illustrated in panels (a)--(c), for all intermediate
$x$ the total formation energy increases with $U$ as in the case of
Li$_x$CoO$_2$ and changes sign from negative to positive for
sufficiently high $U$. For example, for $x=1/2$ the formation energy
is $-0.11$ eV at $U=0$ and increases to +0.21 eV for $U=5.9$ eV. For
$x=1/4$ and $x=3/4$, there are only slightly negative values of around
$-0.02$ eV for $U=0$, which increase to as much as +0.26--0.27 eV as
$U$ increases. We note that the formation energies are higher for this
frozen structure case since the intermediate $x$ species will exhibit
more significant structural relaxations than the endmembers.

For most of the $U$ range, that below the critical $U$ for CO, the DFT
component of the formation energy is approximately constant and does
not strongly influence phase stability. For example, for $x=1/4$ this
component varies by only 25 meV for 0 eV $\le U\le$ 4.7 eV. In
contrast, the $U$--dc component of the formation energy undergoes
significant changes as it increases roughly linearly with $U$; the
change is 0.21 eV over the same $U$ range for $x=1/4$. This $U$--dc
contribution is positive and therefore, as in the case of
Li$_x$CoO$_2$, is what drives the total formation energy towards phase
separation in Li$_x$FePO$_4$ in this regime before CO. One caveat to
this characterization is that the DFT component varies more
considerably (by 90 meV) for Li$_{3/4}$FePO$_4$ in this regime. In
this case it is positive for 2 eV $\le U\le$ 3 eV and thus can be
described as partially responsible for the positive total formation
energy.

Upon CO there is a drastic change in the formation energy components,
as in the case of Li$_x$CoO$_2$. The $U$--dc component drops steeply
to much smaller values of around 0.06--0.08 eV. For $x=1/2$, for
example, the value is 0.33 eV for $U=4$ eV and only 0.07 eV for
$U=4.7$ eV due to the CO transition; the drop in the $U$--dc
contribution is substantial though smaller in magnitude for $x=1/4$
and $x=3/4$. Here as in Li$_x$CoO$_2$ we also find an increase in the
DFT formation energy contribution due to CO. For example, for
Li$_{1/2}$FePO$_4$ the DFT formation energy contribution jumps from
$-135$ meV to +122 meV across the CO phase boundary. Unlike in
Li$_{1/2}$CoO$_2$ [see Fig. \ref{co_frozen_co_decomp}(a)], for
Li$_{1/2}$FePO$_4$ the increase in the DFT component is enough to make
it positive. It is even larger in magnitude than the $U$--dc
component. For $x=1/4$ and $x=3/4$ the DFT component is already
positive but it becomes substantially more positive after CO. Despite
how sharp the changes in formation energy components are, the total
formation energy changes less abruptly from CO. For example, for
$x=1/2$ the total formation energy changes only by 3 meV between $U=4$
eV and $U=4.7$ eV. After the CO transition, the formation energy and
its components are relatively flat versus $U$. For $x=1/2$, for
example, the changes are on the order of only 15--25 meV.


The breakdown of $U$--dc formation energy into filling and ordering
contributions is shown in Fig. \ref{fe_interp_decomps}(d)--(f). As in
Li$_x$CoO$_2$, it can be seen that the average filling contributes
negligibly to the phase separation; the maximum contribution is at
most 25 meV and typically much smaller. Therefore, again the impact of
$U$ and the dc is essentially entirely contained within the ordering
contribution. The ordering contribution tracks the behavior of the
total $U$--dc term and it can be as much as +0.3 eV.

The plots of $N_d$ on each Fe site versus $U$ in Fig.
\ref{fe_interp_decomps}(g)--(i) illustrate that there is also a
$p$--$d$ rehybridization mechanism in Li$_x$FePO$_4$ similar but
smaller than that in Li$_x$CoO$_2$ with differences in $N_d$ around
20--30\% of a full electron between $x=0$ and $x=1$. Like the Fe
magnetic moment data in Fig. \ref{fe_gaps}, these $N_d$ values
illustrate the partial and full CO that occurs as $U$ increases. We
note that without structural relaxations the spread in $N_d$ values
after CO for the intermediate $x$ can be larger than the difference
between $N_d$ of $x=0$ and $x=1$. We find the behavior of $N_d$ with
$x$ is complex and system dependent: in this case $N_d$ for
intermediate $x$ is generally lower than the endmember linear
interpolation, whereas in Li$_x$CoO$_2$ the opposite is true.

\begin{figure}[htbp]
\begin{center}
\includegraphics[width=\linewidth]{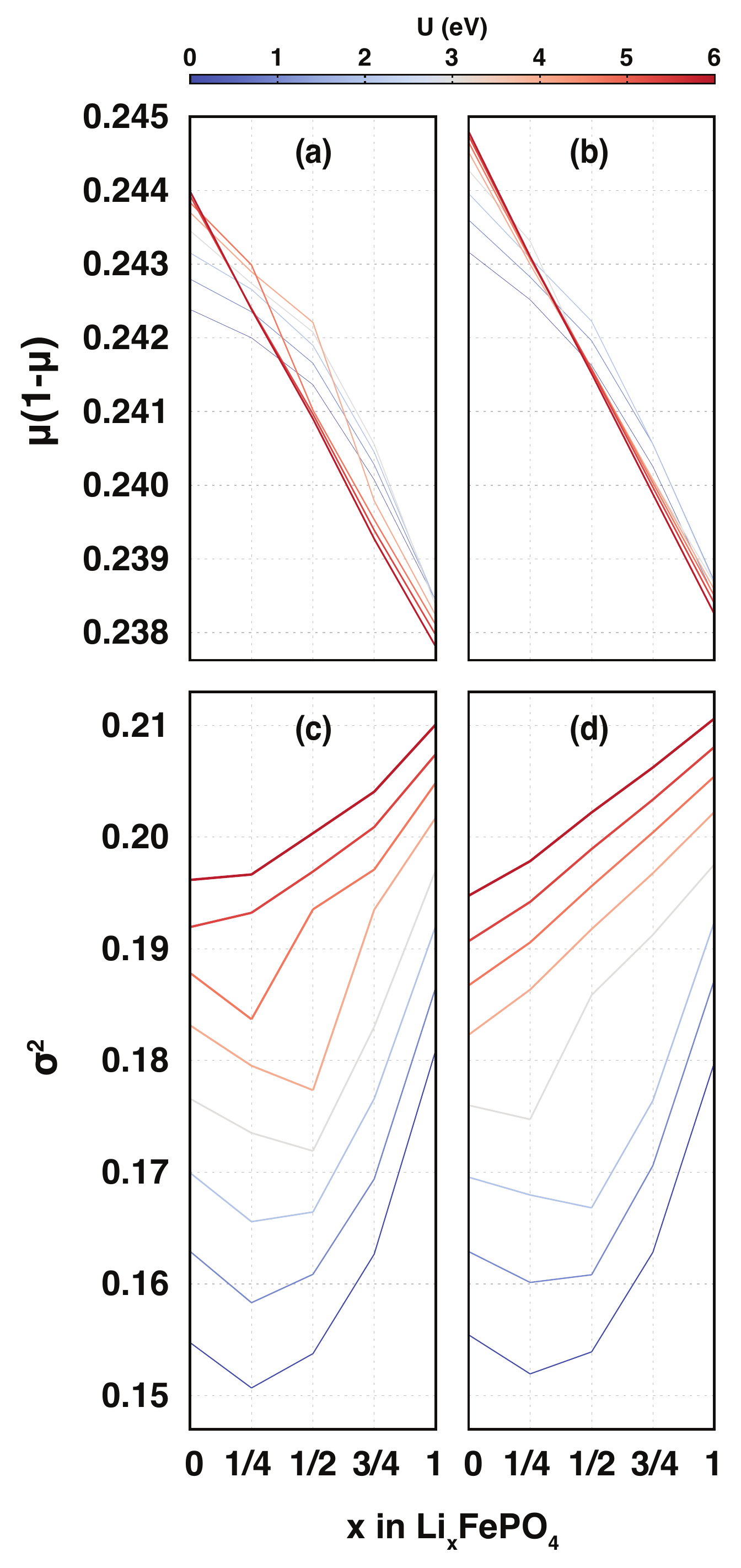}
\end{center}
\caption{Filling factor $\mu(1-\mu)$ as a function of $x$ averaged
  over Fe sites for different $U$ in Li$_x$FePO$_4$ for (a) linearly
  interpolated experimental structures and (b) relaxed structures.
  Panels (c) and (d) show the corresponding plots for the ordering
  factor $\sigma^2$. The line thickness increases for increasing
  values of $U$.\label{fe_mu_sigma}}
\end{figure}

To investigate the origin of the positive $U$--dc phase-separating
contribution to the total formation energy for Li$_x$FePO$_4$, in Fig.
\ref{fe_mu_sigma} we plot the individual $\mu(1-\mu)$ [panel (a)] and
$\sigma^2$ [panel (c)] values as a function of $x$ for different $U$.
For clarity, we take the average over the 4 Fe sites in the primitive
unit cell. As in the case of Li$_x$CoO$_2$, the magnitude of the
filling factor is high (around 0.24) but the changes with respect to
the average of the endmember values are very small (on the order of
$10^{-3}$). This is responsible for the negligible contribution of the
filling component to the total formation energy. As for Li$_x$CoO$_2$,
the filling factor is highest for $x=0$ and $x=1$. Compared to those
of Li$_x$CoO$_2$, the $\mu(1-\mu)$ values of Li$_x$FePO$_4$ are much
nearer to the range expected from nominal electron counting for
Fe$^{2+}$ and Fe$^{3+}$ (0.24--0.25) since there is less hybridization
between $p$ and $d$ states. This is also the reason for the enhanced
magnitude of $\sigma^2$ compared to that of Li$_x$CoO$_2$.

Compared to $\mu(1-\mu)$, $\sigma^2$ has a smaller but still
significant magnitude (around 0.18) for Li$_x$FePO$_4$. The $\sigma^2$
values monotonically increase with $U$ (by around 0.003--0.006 per eV)
for all $x$, which as in the case of Li$_x$CoO$_2$ is expected since
the $U$ and dc terms serve to enhance orbital polarization in the
correlated subspace. Unlike in $\mu(1-\mu)$, there are substantial
deviations of the $\sigma^2$ values for intermediate $x$ compared to
the endmember average. For example, for $x=1/2$ at $U=3$ eV
$\sigma^2$=0.172 as compared to 0.187 for the endmember average. The
ordering factor is consistently lower than the endmember linear
interpolation, thus leading to the positive formation energy
contribution.

One can observe a moderate increase in $\sigma^2$ values for
intermediate $x$ upon CO. For example, for $x=1/2$ $\sigma^2$ jumps
from 0.177 to 0.194 upon CO at $U=4.7$ eV. However, as in the case of
Li$_x$CoO$_2$ the $\sigma^2$ values still are always lower than the
linear interpolation of $x=0$ and $x=1$ values even after CO. For
example, the endmember average $\sigma^2$ is 0.196 for $U=4.7$ eV so
it is still slightly larger than the value for Li$_{1/2}$FePO$_4$.
Therefore, we find that CO alleviates but does not eliminate the
tendency for phase separation with $U$ derived from the ordering of
correlated orbitals.

\begin{figure*}[tb]
\begin{center}
\includegraphics[width=\textwidth]{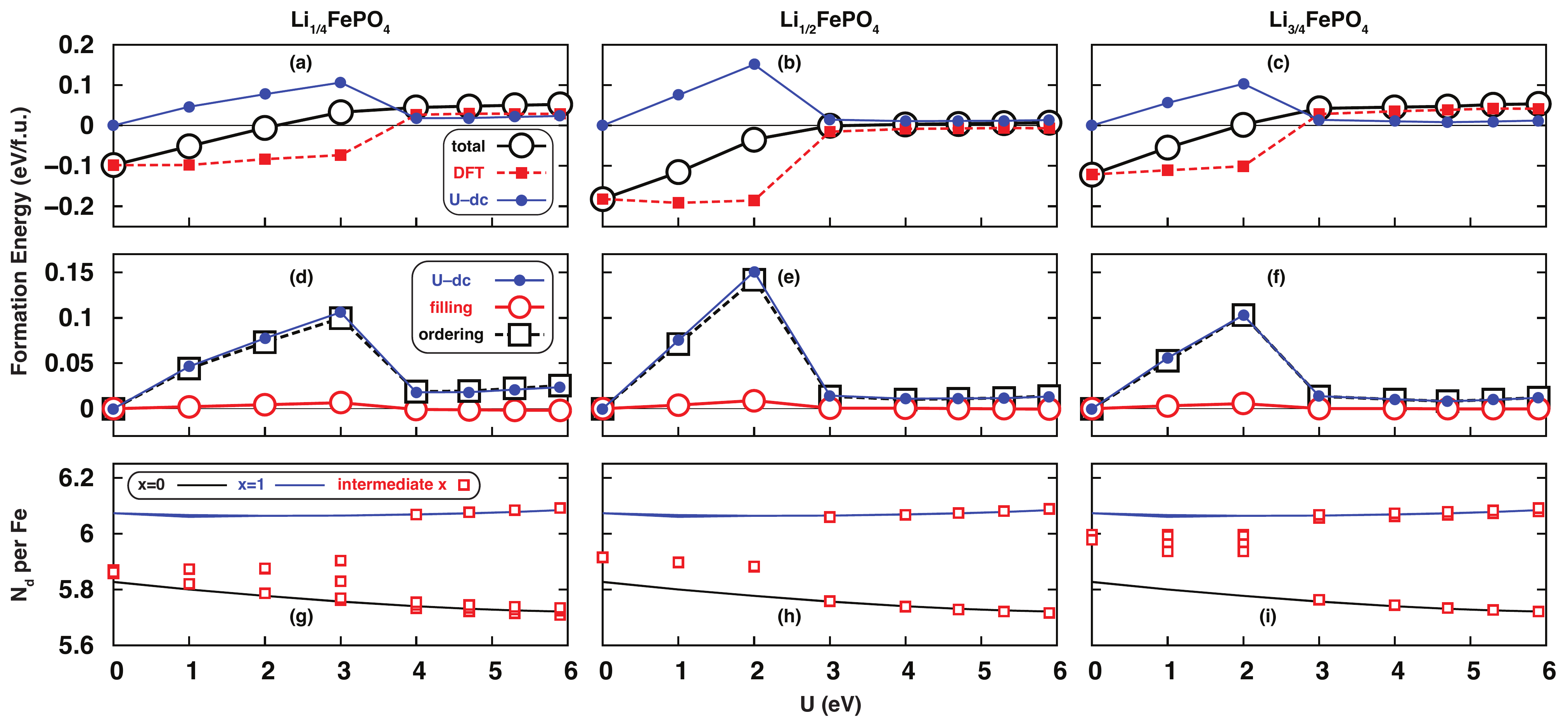}
\end{center}
\caption{Total Li$_x$FePO$_4$ formation energy (open black circles)
  and its DFT (filled red squares) and $U$--dc (filled blue circles)
  components as a function of $U$ for (a) $x=1/4$, (b), $1/2$, and (c)
  $x=3/4$. (d)--(f) show the corresponding plots of Li$_x$FePO$_4$
  $U$--dc (filled blue circles) formation energy component and its
  orbital filling (open red circles) and orbital ordering (open black
  squares) components as a function of $U$. (g)--(i) show the
  corresponding plots of number of $d$ electrons per Fe as a function
  of $U$ for FePO$_4$ (black line), Li$_x$FePO$_4$ (open red squares),
  and LiFePO$_4$ (blue line). All data correspond to the case of
  relaxed structures.\label{fe_rel_decomps}}
\end{figure*}

Figure \ref{fe_rel_decomps}(a)--(c) shows the formation energy
behavior of Li$_x$FePO$_4$ including full structural relaxations. In
this case, since there are more significant total energy lowerings
from relaxing the intermediate $x$ structures compared to those from
relaxing the endmembers, the total formation energy values are
significantly lower. For example, for Li$_{1/4}$FePO$_4$ the maximum
formation energy is 52 meV as opposed to 262 meV in the case of
linearly interpolated endmember experimental structures. Otherwise,
the behavior is generally similar to the case without relaxations. For
all $x$, the total formation energy increases monotonically with $U$
and for sufficiently high $U$ switches from negative to positive. We
note that the formation energy Li$_{1/2}$FePO$_4$ is only slightly
positive (+6.7 meV) for the largest $U$ we considered. This is in
quantitative disagreement with the original DFT+$U$
work\cite{zhou_phase_2004} for reasons which are not clear, but is
consistent with a more recent report.\cite{ong_comparison_2011}

Here again the DFT formation energy contribution is approximately
constant (deviations of at most 25 meV and typically less) before the
CO transition with a slight tendency to increase with $U$. In this
case $x=3/4$ is no longer an exception to the general trend. In
contrast, the $U$--dc contribution is positive and significantly
increases (roughly linearly) with $U$. The largest value it takes on
is 151 meV for $x=1/2$ at $U=2$ eV. Therefore, again it is the
positive $U$--dc contribution that destabilizes the compounds of
intermediate $x$ in this regime.

After CO, the $U$--dc formation energy contribution is significantly
dampened but remains positive. For example, for $x=1/4$ it decreases
from 106 to 18 meV at the CO phase boundary. The effect is similarly
substantial for all $x$. At the same time, the DFT formation energy
component abruptly increases due to CO and becomes positive ($x=1/4$
and $x=3/4$) or much less negative ($x=1/2$). The magnitude of this
increase is more substantial than that of Li$_x$CoO$_2$, which
suggests that in Li$_x$FePO$_4$ CO constitutes a more significant
rearrangement of charge density. The net effect is that upon CO the
increase in total formation energy with $U$ is slowed. For the $x=1/4$
case, for example, the total formation energy increases only by 12 meV
from $U=3$ to $U=4$ eV as opposed to the lower-$U$ regime in which the
same change in $U$ yields increases of around 46 meV. In the CO state
both the DFT and $U$--dc components of the formation energy are nearly
constant with respect to $U$. As such, the total formation energies
saturate to around +50, +6, and +50 meV for $x=1/4$, $x=1/2$, and
$x=3/4$ in the regime of large $U$.

As illustrated in Fig. \ref{fe_rel_decomps}(d)--(f), in this case
again the positive $U$--dc formation energy contribution stems
entirely from the ordering contribution (from the spread in orbital
occupancies); the filling component (from the average orbital
occupancy) is negligible with values less than 10 meV. CO lowers the
filling contribution even further to no more than 1 meV. In contrast,
the ordering contribution is dampened but still positive and
appreciable; it tracks the behavior of the total $U$--dc formation
energy contribution.

The evolution of the $N_d$ values into two discrete groups (Fe$^{3+}$-
and Fe$^{2+}$-like) due to CO is shown in Fig.
\ref{fe_rel_decomps}(g)--(i). A similar magnitude of $p$-$d$
rehybridization is observed based on the overall range of $N_d$. As in
the case without structural relaxations, here for $x=1/4$ and $x=3/4$
though not $x=1/2$ there is a regime of intermediate $x$ in which
there is partial CO in metallic states. After the CO transition the
two groups of $N_d$ values closely match those of $x=0$ and $x=1$,
which suggests the CO is very complete and the local environments of
Fe for intermediate $x$ mimic those of the endmembers.

The filling and ordering factors for relaxed Li$_x$FePO$_4$ are shown
in Fig. \ref{fe_mu_sigma}(b) and \ref{fe_mu_sigma}(d), respectively.
We find similar results when including structural relaxations. The
deviations in $\mu(1-\mu)$ with respect to the linear interpolation of
endmember values are negligible. $\sigma^2$ gradually increases with
$U$ and the values for intermediate $x$ always lag behind the
endmember linear interpolation; the effect is substantially dampened
but not entirely eliminated when CO increases $\sigma^2$ for
intermediate $x$. We note that with structural relaxations after CO
the $\sigma^2$ versus $x$ curves are nearly linear, which illustrates
that relaxations provide a further dampening of the general tendency
towards phase separation with $U$ stemming from the ordering term.

\subsection{Phase stability of Li$_x$CoPO$_4$}\label{lixcopo4}

To further validate our general understanding of the impact of $U$ on
phase stability, we investigate Li$_x$CoPO$_4$. This material is
isostructural to the olivine Li$_x$FePO$_4$ structure shown in Fig.
\ref{structures} with (nominally) an additional electron on the
transition metal site. Li$_x$CoPO$_4$ is of interest since it has been
shown to have a very high voltage (4.8 V) as a cathode
material.\cite{amine_olivine_2000} It is intriguing physically since
unlike Li$_x$FePO$_4$ it does have a stable intermediate compound, for
$x\approx2/3$.\cite{bramnik_phase_2007,ehrenberg_crystal_2009,strobridge_identifying_2014}

We consider the lowest-energy configuration of Li$_{2/3}$CoPO$_4$
deduced by Strobridge \textit{et
  al.}\cite{strobridge_identifying_2014} and study the formation
energy as a function of $U$ with the frozen $U=0$ structures.
Experimentally the Li$_x$CoPO$_4$ system is
AFM,\cite{santoro_magnetic_1966,ehrenberg_crystal_2009} so our
calculations consider the endmembers in the AFM state. We base our
calculations on the magnetic configuration of LiCoPO$_4$, which is
identical to that of
LiFePO$_4$.\cite{santoro_magnetic_1966,santoro_antiferromagnetism_1967}
For Li$_{2/3}$CoPO$_4$ we find the AFM state to be unstable and
devolves into a ferrimagnetic state. In this state one of the Co sites
has no magnetic moment unlike the $\pm\approx2.6-2.7\ \mu_B$ values of
the other 11 leading to a total magnetization of $\approx3.2 \mu_B$.

We find total formation energies of $-0.18$, $-0.14$, and $-0.03$ eV
for $U$ values of 0, 2, and 5.48 eV, respectively. This corresponds to
the same trend of formation energy increasing with $U$. We again find
the positive formation energy contribution is the $U$--dc component,
and here also the filling contribution is very small (magnitude of at
most 20 meV) so the origin is increased ordering of the endmember $d$
states relative to those of the intermediate $x$ species.

For the above analysis we have restricted our attention to the AFM or
AFM-like states as they are the experimental magnetic structure for
Li$_x$CoPO$_4$. However, we note that the magnetic ground state of
Li$_x$CoPO$_4$ predicted by DFT+$U$ changes as a function of $U$,
which is likely the origin of a previous study finding the formation
energy becomes more negative as a function of
$U$.\cite{strobridge_identifying_2014} For LiCoPO$_4$ we find a
ferromagnetic (FM) ground state within DFT with the AFM state 37 meV
higher in energy, whereas for CoPO$_4$ we find a non-spin-polarized
(NSP) ground state with the FM and AFM states 57--61 meV higher in
energy. At $U=2$ the ground state of LiCoPO$_4$ becomes
antiferromagnetic, and by $U=5.5$ eV the ground state for both
endmembers is AFM consistent with experiments. For $x=2/3$ we find the
FM state is unstable within DFT and the NSP state is 0.54 eV higher in
energy than the ferrimagnetic state. At $U=2$ eV the FM state becomes
metastable only 7 meV above the ferrimagnetic state, and for $U=5.5$
eV the FM state becomes the ground state with the ferrimagnetic state
68 meV high in energy. In any case, DFT+$U$ does predict the $x=2/3$
phase to be stable, qualitatively consistent with experiment.

\subsection{Impact of double counting on phase separation trend}\label{strawman}

Our preceding analysis shows that it is valuable to recast the $U$--dc
energy as the sum of ordering and filling terms, as the filling term
was shown to have negligible impact on formation energies. However,
there is still utility in analyzing what can be learned by separately
inspecting $E_U$ and $E_{\mathrm{dc}}$, given that the $E_U$ term
specifically is handled much more precisely in the context of
DFT+DMFT.

The simplest possible explanation for the positive $U$--dc formation
energy contribution comes from the form of the dc term Eq.
\ref{flldc}. We expect the Li in Li$_x$CoO$_2$ and other intercalation
materials will be ionized and donate some amount of charge (depending
on the degree of hybridization with O $p$ states) to Co. Therefore, we
might make the assumption that $N_d$ is linear in $x$. Since energy
contributions linear in $x$ do not contribute to the formation energy
by the definition of Eq. \ref{fe}, this means the term linear in $N_d$
in the dc term cannot contribute to the formation energy. The other
part of the dc, however, yields a term proportional to $-N_d^2$ (a
negative quadratic in $N_d$) in the total energy. With the assumption
of a linear relationship between $N_d$ and $x$, this term is a
negative quadratic in $x$ that necessarily provides a positive
formation energy contribution.

This simple kind of argument fails to fully describe the observed
behavior for the following reasons, which we illustrate using the
simplest example of Li$_{1/2}$CoO$_2$ without CO considering the
frozen $U=0$ structures. The first two reasons come from the fact that
the variation of $N_d$ with $x$ is in fact substantially nonlinear, as
can be easily seen for the $U>1$ eV data in Fig.
\ref{co_frozen_decomp}(c). In this regime, $N_d$ for $x=1/2$ is much
closer to the $x=1$ value than the $x=0$ value. For example, for $U=5$
eV the $N_d$ values are 7.19, 7.28, and 7.30 for $x=0$, $x=1/2$, and
$x=1$, respectively. First of all, this means that the contribution of
the dc energy that is proportional to $N_d$ will contribute to the
formation energy. For example, for this case it takes on the value of
+89 meV for $U=5$ eV.

Secondly, the quadratic part of the dc energy no longer necessarily
gives a positive formation energy contribution for $N_d$ that is
nonlinear in $x$. One can parametrize the deviation from linearity
using the additive form $N_d(x)=\overline{N_d(x)}+\delta_x$, where
$N_d(x)$ is the $N_d$ value for Li concentration $x$,
$\overline{N_d(x)}=(1-x)N_d(0)+xN_d(1)$ is the linear value of
$N_d(x)$, and $\delta_x$ is the deviation from linearity. Here we
assume a single site for clarity. It then can be shown that the
formation energy contribution stemming from the quadratic part of the
dc energy
is \begin{align}FE_{\mathrm{dc}}^{\mathrm{quad}}(x)=\frac{1}{2}U\bigg\{&x(1-x)\left[N_d(1)-N_d(0)\right]^2\nonumber\\&-2\overline{N_d(x)}\delta_x\left(1+\frac{\delta_x}{2\overline{N_d(x)}}\right)\bigg\}.\end{align}
Note that the term proportional to $\delta_x^2$ can be ignored as
$\delta_x$ should be much smaller than $2\overline{N_d(x)}$. When
$\delta_x=0$ ($N_d$ is linear with $x$), we indeed have a positive
formation energy contribution since the second term vanishes and the
first term will be positive since $N_d(0)\ne N_d(1)$. However, this
second term can lead to a significant negative (phase stabilizing)
formation energy contribution when $\delta_x$ is positive and
non-negligible. This actually results in a strongly negative formation
energy contribution in the present case: for $U=5$ eV, for example,
the formation energy contribution from the quadratic part of the dc is
$-1.29$ eV. Therefore, the dc term is not responsible for driving
phase separation in this regime of $U$.

The third and final reason the simple argument fails is that the $E_U$
term also contributes significantly to the formation energy. For
example, for $U=5$ eV the formation energy contribution from the
interaction term is +1.78 eV, strongly driving phase separation.

Additionally, even in the limit of small $U$, in which the dc does
provide a phase separation contribution, it is not fully responsible
for the overall trend. In this limit, the impact of $U$ on the
interaction and dc formation energy components depends solely on the
DFT density matrix. Computing $\partial E_U/\partial U$ and $-\partial
E_{dc}/\partial U$ from Eqns. \ref{dftu_u_eqn} and \ref{flldc},
respectively, and evaluating them at $U=0$, we find that the dc energy
contributes $7\times10^{-3}$ and the interaction energy contributes a
larger $13\times10^{-3}$ to the change in total formation energy.

\subsection{Average intercalation voltage of Li$_x$CoO$_2$ and Li$_x$FePO$_4$}\label{voltage}

\begin{figure}[htbp]
\begin{center}
\includegraphics[width=\linewidth]{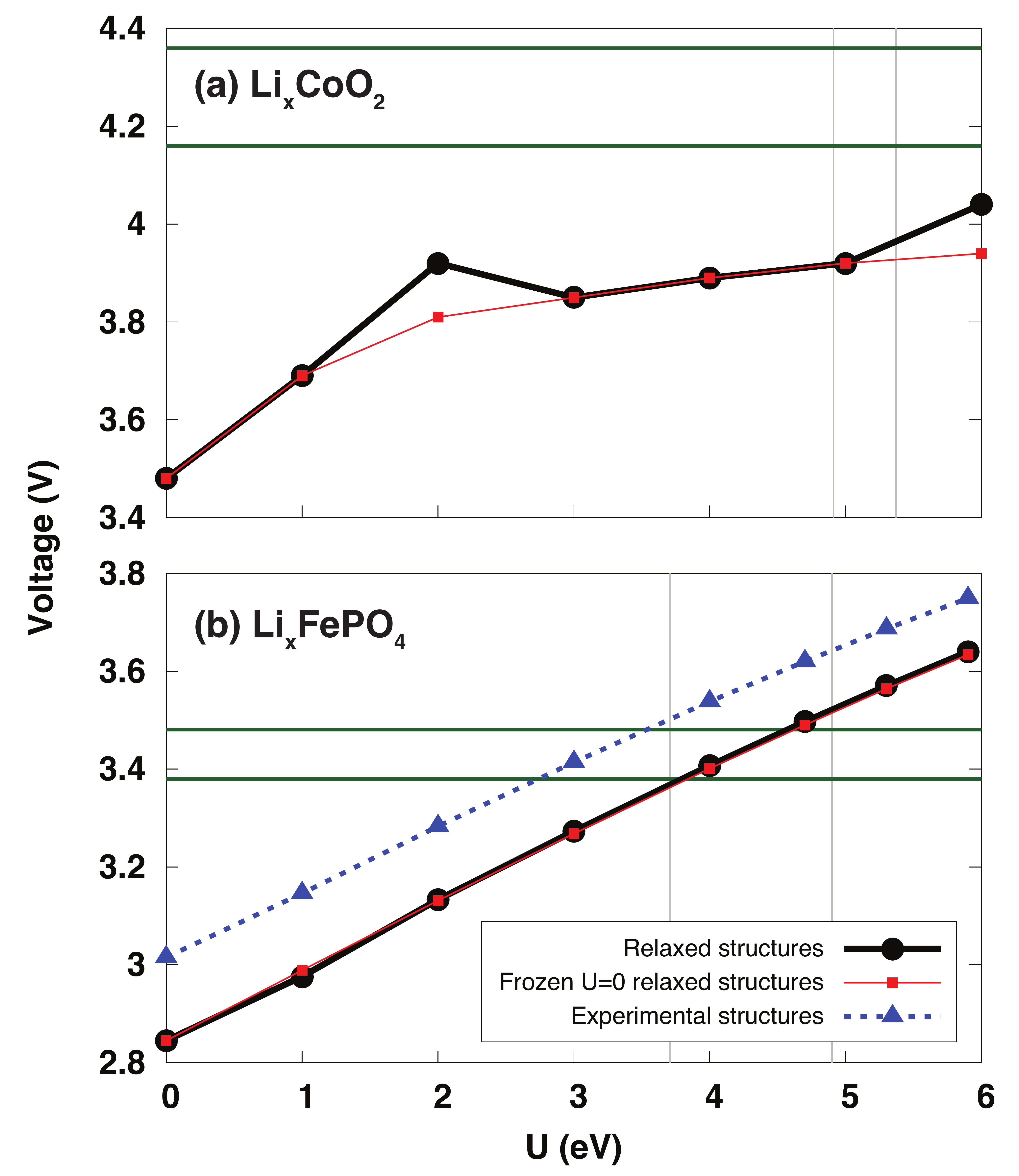}
\end{center}
\caption{Average intercalation voltage of (a) Li$_x$CoO$_2$ and (b)
  Li$_x$FePO$_4$ as a function of $U$ shown using relaxed structures
  and frozen $U=0$ relaxed structures. For Li$_x$FePO$_4$ the results
  using the experimental structures are also shown. Green lines
  indicate the range of the average voltage measured from experiment,
  and grey lines indicate $U$ values of the endmembers computed via
  the linear response
  method.\cite{zhou_first-principles_2004}\label{voltages}}
\end{figure}

Up to now our analysis of the DFT+$U$ energetics has focused on the
formation energy, and we seek other observables to probe the fidelity
of our total energy calculations. Therefore, we now turn to average
intercalation voltage, which is a function of the energy difference of
the cathode endmembers and the energy of bulk Li (see Eq.
\ref{eq:voltage}) and can be measured experimentally.

Figure \ref{voltages} shows the behavior of the average intercalation
voltage as a function of $U$. As found previously, the voltages tend
to increase with $U$ for both Li$_x$CoO$_2$ and
Li$_x$FePO$_4$.\cite{zhou_first-principles_2004} For Li$_x$CoO$_2$
using relaxed structures, the computed voltage exhibits one
discontinuity at $U=2$ eV after which CoO$_2$ gaps and another at
$U=5$ eV after which LiCoO$_2$ becomes high spin. Otherwise, the
values are almost identical to those found using the frozen $U=0$
structures. The DFT value is 3.48 V and there is an increase to 3.92 V
at $U=5$ eV. We note that the predicted voltages for Li$_x$CoO$_2$ are
smaller than those reported in the work of Zhou \textit{et
  al.},\cite{zhou_first-principles_2004} but agree with several more
recent
studies.\cite{chevrier_hybrid_2010,aykol_local_2014,aykol_van_2015}
Ultimately, DFT+$U$ underpredicts the average voltage of Li$_x$CoO$_2$
compared to the experimental value of 4.26 V.\cite{amatucci_coo2_1996}
For Li$_x$FePO$_4$ using relaxed structures, the DFT voltage is 2.85 V
and increases approximately linearly to 3.50 V at $U=4.7$ eV, in
agreement with the experimental voltage of 3.43
V.\cite{padhi_effect_1997,yamada_optimized_2001} Using the
experimental structures, the predicted intercalation voltage is
enhanced by 0.1-0.2 V.

In Li$_x$FePO$_4$, the two endmembers have similar properties as they
are AFM insulators with very well localized electronic states.
Therefore, it is possible that DFT+$U$ is effective at describing the
energetics of both phases and hence can capture the voltage
effectively. However, in Li$_x$CoO$_2$ one endmember (LiCoO$_2$) is a
band insulator while the other is a Fermi liquid (in experiment); the
latter is found to be a magnetic metal or insulator within DFT+$U$
depending on $U$. We speculate that the distinct nature of the
endmembers for Li$_x$CoO$_2$, and their inadequate description within
DFT+$U$, will lead DFT+$U$ to make energetic errors which vary
substantially for the different endmembers causing a worse description
of the voltage as compared to the olivines. A forthcoming DFT+DMFT
study will address this issue.

\subsection{Li order-disorder transition temperature of Li$_{1/2}$CoO$_2$}\label{od_temp}

Another observable with which we can assess the accuracy of DFT+$U$
energetics, albeit at a fixed composition, is the order-disorder (O-D)
transition temperature for Li. This is the temperature above which the
Li ions and vacancies become disordered. Here we consider
Li$_{1/2}$CoO$_2$, whose experimental O-D temperature is 333
K.\cite{reimers_electrochemical_1992}

We estimate the O-D transition temperature based on the ground state
total energies of the ordered and disordered (SQS) phases using the
equation $$T_{\mathrm{O-D}}=\frac{E_{D}-E_{O}}{k_B\ln(2)},$$ where
$k_B$ is the Boltzmann constant and the factor of $\ln(2)$ comes from
the entropy of mixing for $x=1/2$. This expression is an approximation
to performing finite-temperature Monte Carlo simulations on a cluster
expansion based on such ground-state total energies. Past work
suggests that this simpler expression should reasonably capture trends
in
$T_{\mathrm{O-D}}$.\cite{lu_first-principles_1994,jiang_first-principles_2004,jiang_first-principles_2005,jiang_first-principles_2009}
Performing a cluster expansion of Li$_{x}$CoO$_2$ using DFT+$U$ would
likely be formidable given the many spurious ordered states which form
at nonstochiometric compositions, as will be elucidated below.

\begin{figure}[htbp]
\begin{center}
\includegraphics[width=\linewidth]{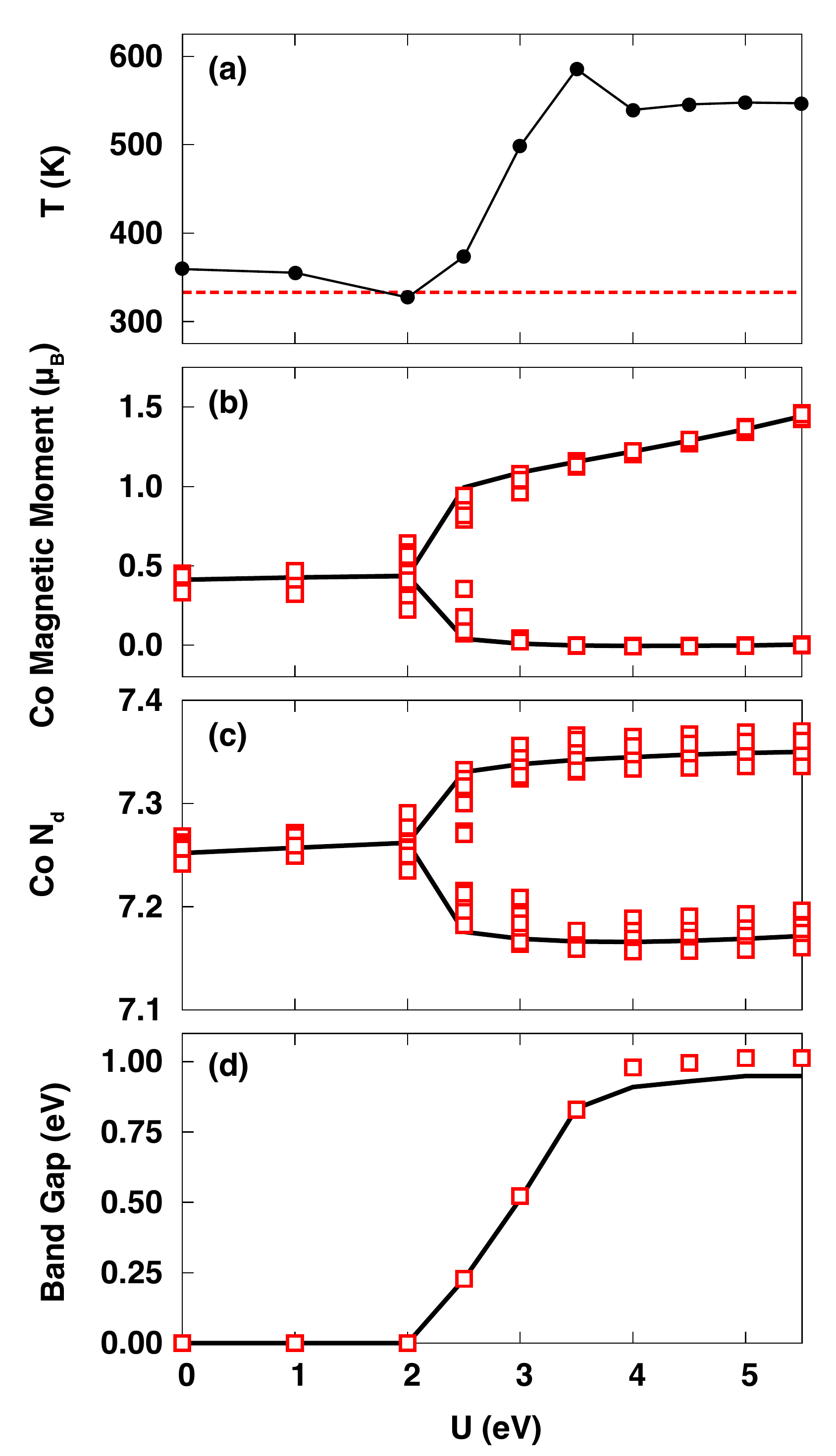}
\end{center}
\caption{(a) Li order-disorder transition temperature, (b) Co magnetic
  moments, (c) Co N$_d$, and (d) band gap for Li$_{1/2}$CoO$_2$ as a
  function of $U$ using the frozen $U=$ 0 eV structures. The
  disordered phase is modeled by an optimal 42-ion special quasirandom
  structure. In panel (a) the horizontal red line indicates the
  experimental order-disorder transition temperature from Ref.
  \citenum{reimers_electrochemical_1992}. For the other panels black
  lines (red squares) correspond to results from the ordered
  (disordered) structure.\label{sqs}}
\end{figure}

Our predicted $T_{\mathrm{O-D}}$ for an optimal 42-ion SQS cell frozen
to the relaxed $U=0$ structures is illustrated in Fig. \ref{sqs}(a).
Here within DFT the temperature is overestimated by around 25 K with
respect to the experimental value. For small values of $U$ the
temperature decreases by as much as 32 K, bringing the prediction
closer to the experimental value. However, for $U>2$ eV, there is a
very rapid increase in the predicted transition temperature to values
of around 500--600 K, nearly a factor of two greater than the
experimental value.

The Co magnetic moments and $N_d$ values as a function of $U$ are
shown in panels (b) and (c) of Fig. \ref{sqs}, respectively. For small
$U$ there are small deviations of the magnetic moments and $N_d$
compared to those of the ordered structure due to the different local
environments of the Co in the disordered phase. The range of these
values are enhanced for $U=2$ eV, above which both the ordered and
disordered cells charge order and open an electronic band gap [see
  panel (d)]. After the CO transition, there is still some spread in
the magnetic moment and $N_d$ values for the disordered phase with
respect to those of the ordered phase. In this particular case,
structural relaxations actually amplify these trends, resulting in a
substantially higher $T_{\mathrm{O-D}}$ for $U=4$ eV, for example (see
Supplementary Material for more details).

This demonstrates that the CO transition found within the DFT+$U$
approach is largely responsible for the erratic behavior of the
predicted Li order-disorder transition temperature for
Li$_{1/2}$CoO$_2$. In the Supplementary Material, we show results for
additional SQS cells with and without structural relaxations. We find
that within DFT $T_{\mathrm{O-D}}$ can be overestimated or
underestimated depending on the particular cell. A substantial
increase in $T_{\mathrm{O-D}}$ due to CO is often observed, though it
is not always as dramatic this particular case and does not always
lead to as large a disagreement with the experimental value. In
general, the results indicate that DFT+$U$ is highly unreliable in
predicting the order-disorder transition temperature due to CO. This
further supports the notion that CO is an artifact of the approximate
nature of the interaction in DFT+$U$, consistent with the fact that it
is not observed in experiment. While spurious CO appears to be
essential for properly attaining a phase stable system in
Li$_{1/2}$CoO$_2$, as discussed in Sec. \ref{charge_ordering}, it
appears to cause severe problems in the SQS supercell. While it is
possible that more reasonable behavior emerges for substantially
larger SQS supercells, which would be computationally expensive, this
is beyond the scope of the current paper.


It should be noted that the spurious behavior we have documented in
the context of this SQS calculation of $T_{\mathrm{O-D}}$ would likely
manifest itself in any parameterization of the cluster expansion. In
the best case scenario, this could lead to a complicated cluster
expansion with long range interactions. The entire procedure could
become ill-posed in the worst case with numerous orbitally ordered
and/or CO states close in energy for a given Li ordering. This
behavior does not bode well for the application of DFT+$U$ in alloy
thermodynamics. Given that DFT+DMFT does not necessitate
charge/orbital ordering in order to capture the energetics of strongly
correlated systems, it is likely not to share these deficiencies.

\section{Conclusions}\label{conclusions}

At the DFT+$U$ level of theory, the on-site $U$ tends to destabilize
intermediate-$x$ compounds of both phase stable Li$_x$CoO$_2$ and
phase separating Li$_x$FePO$_4$, though qualitatively correct results
are obtained for formation energies in both cases for physically
reasonable values of $U$. A new formation energy decomposition, which
disentangles the distinct roles of the filling and ordering of $d$
orbitals, reveals that reduced orbital ordering in compounds of
intermediate Li concentration relative to the endmembers is
responsible for this effect. For intermediate $x$ DFT+$U$ predicts
charge ordering which opens electronic band gaps and dampens, though
does not eliminate, the tendency for phase separation. While charge
ordering and an appreciable band gap are spurious for
Li$_{1/2}$CoO$_2$, charge ordering is essential for retaining a phase
stable system, as found in experiment, at reasonable values of $U$.
This deficiency appears to be a manifestation of the Hartree-Fock
treatment of the interaction energy within DFT+$U$, whereby static
ordering is required to properly capture the energetics of strong
electronic correlations. Indeed, it is demonstrated that the origin of
the charge ordering arises from interaction energy within DFT+$U$, as
opposed to the double counting correction. Structural relaxations,
which similarly only reduce the magnitude of the $U$-induced
destabilization of intermediate compositions, have a more significant
impact on Li$_x$FePO$_4$ than Li$_x$CoO$_2$ due to the stronger
electron-lattice coupling. The same generic formation energy behavior
is observed in Li$_x$CoPO$_4$, for which experiment dictates that a
stable intermediate-$x$ compound exists unlike for isostructural
Li$_x$FePO$_4$.

The Li order-disorder transition temperature of Li$_{1/2}$CoO$_2$
behaves erratically when computed using SQS supercells within DFT+$U$
as a result of charge ordering; severe overestimations are not
uncommon. Along with a lack of experimental evidence and the induction
of a spurious insulating state, this further supports the notion that
static charge ordering is likely an unphysical artifact of the method.
Future work should focus on improved methods for predicting total
energies in realistic correlated electron materials using dynamical
mean-field theory or other approaches. Given that many of the
deficiencies that we have identified in this study appear to be
associated with the Hartree-Fock treatment of the interaction energy
within DFT+$U$, it is likely that DFT+DMFT could offer substantial
improvements.

\begin{acknowledgments}
This research used resources of the National Energy Research
Scientific Computing Center, a DOE Office of Science User Facility
supported by the Office of Science of the U.S. Department of Energy
under Contract No. DE-AC02-05CH11231. E.B.I. gratefully acknowledges
support from the U.S. Department of Energy Computational Science
Graduate Fellowship (Grant No. DE-FG02-97ER25308).
\end{acknowledgments}

\bibliography{main}

\end{document}